\documentclass[11pt]{article}
\usepackage[letterpaper,margin=1in]{geometry}
\usepackage[T1]{fontenc}
\usepackage{lmodern,amsmath,amssymb,amsthm,mathtools,microtype}
\usepackage{enumitem,booktabs,array,xcolor,xurl}
\usepackage{algorithm,algpseudocode}
\usepackage{placeins}
\usepackage{tikz}
\usetikzlibrary{arrows.meta,positioning,calc,fit,backgrounds}
\usepackage[colorlinks=true,linkcolor=blue!45!black,citecolor=blue!45!black,urlcolor=blue!45!black]{hyperref}
\hypersetup{pdftitle={A Polynomial-Time Sixth-Power Excluded Grid Theorem},pdfauthor={Mohammad Hajiaghayi}}
\setlist{itemsep=2pt,topsep=4pt}
\newtheorem{theorem}{Theorem}[section]
\newtheorem{lemma}[theorem]{Lemma}
\newtheorem{proposition}[theorem]{Proposition}
\newtheorem{corollary}[theorem]{Corollary}
\theoremstyle{definition}
\newtheorem{definition}[theorem]{Definition}

\theoremstyle{remark}

\newcommand{\tw}{\operatorname{tw}}

\newcommand{\polylog}{\operatorname{polylog}}
\newcommand{\tO}{\widetilde O}
\newcommand{\Grid}{\mathsf{Grid}}

\newcommand{\calR}{\mathcal R}

\title{A Polynomial-Time Sixth-Power Excluded Grid Theorem}
\author{Mohammad Hajiaghayi\thanks{Department of Computer Science, University of Maryland, College Park. Email: \href{mailto:hajiagha@umd.edu}{\texttt{hajiagha@umd.edu}}.}}
\date{November 12, 2026}
\begin{document}
\maketitle

\begin{abstract}
We give a polynomial-time construction for a sixth-power excluded-grid theorem. Every graph whose treewidth is at least $g^6$ times a fixed polylogarithmic factor contains a $g\times g$ grid minor that our randomized algorithm finds with constant probability. The algorithm outputs an explicit minor model, and the exponent in its running time is independent of $g$. This improves the ninth-power existence bound of Chuzhoy and Tan~\cite{CT21} and the $g^{98}$ threshold in the polynomial-time construction of Chekuri and Chuzhoy~\cite{CC16}. Quantitative bounds relating treewidth to grid size are central to bidimensionality, developed by Demaine, Hajiaghayi, and collaborators~\cite{DH05,DH08}.

The main ingredient is an asymmetric local routing theorem that uses substantially fewer attachment paths than backbone paths. A balanced recursion preserves connectivity by charging deletions to the boundary of a fixed reference cluster. We construct the smaller attachment sets through a routing hierarchy: either it exposes an accessible well-linked boundary, or its quotient graphs contain the desired grid. These arguments implement cluster splitting using approximate cuts and flow computations. An explicit cut-matching construction and a constructive grid-extraction procedure complete the algorithm.

We also give a general packing--covering hypothesis for trees meeting three prescribed terminal sets under which the exponent can be reduced below six. A companion Lean development verifies the new construction, its success guarantee, and the composition of its polynomial running time, using explicit interfaces for the published algorithms and standard finite-data operations.
\end{abstract}

\section{Introduction}
\label{sec:intro}

How much treewidth forces a large grid minor? Let $f(g)$ be the least integer such that every finite graph of treewidth at least $f(g)$ contains the $g\times g$ grid. Robertson and Seymour proved that $f(g)$ is finite~\cite{RS86}. Determining its growth has since become a basic quantitative problem in graph minor theory. A grid gives a particularly concrete certificate of large treewidth: it exposes many disjoint paths, a two-dimensional pattern of intersections, and a common structure on which a wide range of graph parameters are large.

This relation is also central to bidimensionality. A minor-monotone parameter that is quadratic on a grid turns a bound on grid size into a bound on treewidth, after which dynamic programming becomes applicable. The work of Demaine and Hajiaghayi on bidimensionality and approximation schemes~\cite{DH05}, and on linear grid-minor bounds in fixed-minor-free classes~\cite{DH08}, made this connection a general algorithmic tool; see also~\cite{DFHT05,DHK09}. The unrestricted excluded-grid problem is quantitatively different. For every fixed excluded minor, the treewidth/grid relation is already linear. The present paper concerns arbitrary graphs.

\paragraph{The quantitative gap.}
Robertson, Seymour, and Thomas proved the lower bound $f(g)=\Omega(g^2\log g)$ using graphs of large girth, and suggested that a matching upper bound might hold~\cite{RST94}. Thus the near-quadratic lower bound is established; its sufficiency is conjectural. Demaine and Hajiaghayi proposed a cubic upper bound, together with a cubic tightness conjecture, in~\cite[Conjecture~16]{DH05Map}; the later journal treatment records the conjecture as well~\cite[Conjecture~39]{DHK09}. These are distinct predictions for the eventual answer.

For many years, the upper bounds were exponential or worse~\cite{RST94,LS15}. Chekuri and Chuzhoy obtained the first polynomial bound, with exponent $98$, together with a randomized polynomial-time construction~\cite{CC16}. Chuzhoy reduced the exponent to $36$ and then $19$~\cite{Chuzhoy15,Chuzhoy16}. Chuzhoy and Tan established the ninth-power theorem through a local crossbar argument~\cite{CT21}. A registered 2026 submission to the Lax Lean Archive reports an exponent-eight existential bound and expressly omits running-time and probability guarantees~\cite{Bonnet26}.

For constructive applications, the dependence of the running time on $g$ is a separate requirement. A running time $F(g)n^{O(1)}$ treats the grid size as a fixed parameter; a bound of $(n+m+g)^{O(1)}$ also permits it to grow with the input while preserving polynomial running time. This distinction matters when grid extraction is used inside an approximation or routing algorithm. Our theorem gives the uniform polynomial guarantee at a sixth-power threshold.

\begin{theorem}[Polynomial-time sixth-power excluded grid theorem]\label{thm:main}
There are absolute constants $C,c>0$ and a randomized algorithm with the following guarantee. Given a graph $G$ and an integer $g\ge2$, the algorithm runs in
\[
 (|V(G)|+|E(G)|+g)^{O(1)}
\]
time, with an absolute exponent. If
\[
 \tw(G)\ge Cg^6\bigl(\log_2(g+2)\bigr)^c,
\]
it returns an explicit $g\times g$ grid-minor model with probability at least $2/3$. Every returned model is verified in $G$. Independent repetition until a verified model is found has expected polynomial running time on inputs satisfying the treewidth hypothesis.
\end{theorem}

The polynomial improvement is the main point of the theorem. We do not optimize the logarithmic exponent. In the inverse formulation, every graph of treewidth $k$ contains a square grid of side $\widetilde\Omega(k^{1/6})$. Neither formulation gives a matching bound: the general lower bound remains $\Omega(g^2\log g)$.

\paragraph{The local theorem.}
An $(A,B,X)$-crossbar of width $\rho$ consists of $\rho$ vertex-disjoint $A$--$B$ paths, together with one path from each of them to a distinct vertex of $X$. The latter paths are disjoint, and each meets the former paths only at its own attachment vertex. A strong path-of-sets system of width $w$ and length $\ell$ consists of $\ell$ disjoint connected clusters with linked, node-well-linked interfaces of size $w$, joined in order by disjoint linkages; the full definition appears in Section~\ref{sec:prelim}.

\begin{theorem}[Constructive local dichotomy]\label{thm:local}
There are absolute constants $C_0,c_0>0$ with the following property. Let $w,\rho,\ell\ge2$ be integers. Suppose $A,B,X$ are disjoint vertex sets of common size $\kappa$ in a graph $H$, every vertex of $X$ has degree one, and there are linkages of size $\kappa$ from $A$ to $B$ and from $A$ to $X$. If
\[
 \kappa\ge C_0 w\rho\ell\bigl(\log_2(w\rho\ell+2)\bigr)^{c_0},
\]
then a polynomial-time algorithm finds either an $(A,B,X)$-crossbar of width $\rho$, or a minor of $H$ of maximum degree four containing a strong path-of-sets system of width $w$ and length $\ell$.
\end{theorem}

The running time is polynomial in the number of vertices and edges of $H$ and the displayed integer parameters. The linkages can be supplied or found by maximum flow. This algorithmic statement uses the constructive form of well-linkedness boosting, with its additional logarithmic loss; it does not assume an exact sparsest-cut oracle.

\paragraph{Unequal interfaces and a polynomial supply step.}
For the general construction we prove the asymmetric version in Theorem~\ref{thm:unbalanced-local}. The two backbone interfaces need size $\widetilde O(w\rho\ell)$, whereas the attachment interface needs only $\widetilde O(w\rho)$. With $w=\rho=\ell=q^2$, these are sixth-power and fourth-power quantities. A strong path-of-sets system returned locally gives a grid of side $\Omega(q)$. Otherwise crossbars assemble, by an explicit $O(\log^2q)$-round cut player, into a grid of side $\Omega(q/\log^{5/2}q)$.

The smaller attachment requirement is what makes the global supply step polynomial. Inside each odd cluster of a constructive strong path-of-sets system, a routing hierarchy either exposes an accessible boundary of the required smaller size or produces a grid in a quotient. We trim the complement, preserving the two large interfaces, and use constructive well-linkedness boosting with unequal terminal classes. The even clusters remain available for stitching. This proves the general-graph construction using polynomial flows, approximate cuts, and explicitly represented minor maps.

Our proof uses the constructive path-of-sets supplier of Chekuri and Chuzhoy~\cite{CC16} and their degree-three treewidth sparsifier~\cite{CC15Sparsifiers}. Work of Hayashi, Kawarabayashi, and Kobayashi~\cite{HKK26} gives sharper bounds on minimal pairs of linkages and the size of treewidth sparsifiers. The polynomial-size sparsifier of~\cite{CC15Sparsifiers} already suffices for the logarithmic normalization used here.

\paragraph{Routes to smaller exponents.}
For three terminal sets, call a connected subgraph \emph{colorful} if it meets all three. Write $\nu$ for the largest number of disjoint colorful subgraphs and $\tau$ for the smallest vertex set meeting all of them. We prove that a bound
\[
 \tau\le Cg^a\nu\bigl(\log_2(g\nu+2)\bigr)^b
 \quad\text{in every $g\times g$-grid-free graph}
\]
would imply $f(g)=\tO(g^{a+2})$. Every decrease in $a$ gives the same decrease in the resulting grid exponent; any fixed $a<4$ would improve the sixth-power bound. The linear case $a=1$ gives the cubic exponent, up to logarithmic factors. The reduction isolates the remaining polynomial loss in one covering hypothesis. It establishes an implication and does not prove that hypothesis.

\paragraph{Organization.}
Section~\ref{sec:approach} gives the proof overview. Sections~\ref{sec:source}--\ref{sec:ordered} develop the local consumer, and Section~\ref{sec:unbalanced-local} proves the unequal-interface theorem. Sections~\ref{sec:weighted-separators}--\ref{sec:asymmetric-split} construct the smaller hairs. Section~\ref{sec:polynomial-assembly} gives the certified assembly. We collect the local and global parameter choices in Section~\ref{sec:parameters}; Section~\ref{sec:conditional-cover} explains the general conditional improvement in the exponent.

\section{Our approach and techniques}
\label{sec:approach}

\subsection{Two different width requirements}

We build a strong path-of-sets chain and use its odd clusters to realize successive matchings on a common set of backbone paths. A local call either supplies the required crossbar or already contains a grid. The difficulty is to provide the attachments used by those crossbars without spending another power of the grid size.

The local theorem separates two requirements. To obtain a strong system of width $w$ and length $\ell$, or a crossbar of order $\rho$, the backbone linkage needs $\widetilde O(w\rho\ell)$ paths, but the attachment linkage needs only $\widetilde O(w\rho)$. Taking $w=\rho=\ell=q^2$ gives a sixth-power backbone and a fourth-power attachment set. A polynomial construction of the smaller set will suffice.

Here is why the length need not enter both widths. Fix the large $A$--$B$ linkage and repeatedly try to connect its available paths to $X$. A flow of value $\rho$ gives a crossbar. Otherwise a separator of size less than $\rho$ blocks access; remove the backbone paths represented in that separator and continue. Only afterwards choose an $X$--$A$ linkage ending on the remaining backbone paths. Each of its paths must cross every recorded separator. Apart from fewer than $\rho$ exceptions per layer, it meets one of the removed backbone paths. Counting these contacts layer by layer gives many incidences between the two families.

We now reverse their roles. The smaller linkage becomes the row family, while the removed backbone paths supply auxiliary segments. Choose the smaller linkage by one minimum-cost flow, giving cost zero to the fixed backbone and cost one elsewhere. In the union of the two families, every positive-cost edge is indispensable. Contracting their common edges and the off-row portions of the backbone therefore makes the rows a perfect unique unpaired linkage. Segment the backbone images \emph{after} these contractions. The incidence count supplies enough disjoint segments meeting many distinct rows. No common set of paths has to avoid the exceptional vertices at every layer.

The same procedure proves both local theorems. It only needs to retarget the smaller linkage to the surviving half of $A$. In the balanced case, restrict the given full linkage to those endpoints. In the unequal-interface case, node-well-linkedness of $A$ supplies a routing with congestion two, which becomes a disjoint linkage after losing a factor of two in its size. This short retargeting argument is the only difference between the two inputs.

\subsection{The local consumer}

The resulting core has at most $O(H\rho)$ unique rows and many disjoint auxiliary segments, each meeting $64H$ distinct rows. An auxiliary segment may visit the rows in an arbitrary order, so it need not resemble a grid column. We use it to provide well-linked terminal mass; the unique rows provide order.

Mark contacts and group consecutive contacts into blocks along the rows. Pruning the block--segment incidence graph and decomposing at sparse terminal cuts gives disjoint clusters with fully well-linked terminal sets. The total terminal mass is proportional to the number of auxiliary segments. The guarantee holds for every terminal cut, with no cap on its smaller side.

Uniqueness lets us sweep one frontier vertex along each row. If every unfinished row were blocked by an edge going forward to another row, a cycle of dependencies would permit an exchange of row suffixes, contradicting uniqueness. A frontier therefore exists that balances any specified terminal mass. Recursively sweep the current cells. For each cluster, try to route a small fraction of its terminals to the frontier. A successful flow captures ports on distinct rows. A failed flow gives a small separator, behind which one component contains most of the terminals.

Restricting to that component may destroy well-linkedness. Repeating an approximate decomposition at every level would compound the loss. Instead keep the original cluster $C_0$ and terminal set $\Gamma_0$ fixed. In the current vertex set $U$, inject one unit for every edge leaving $U$ in $C_0$, give internal edges capacity one, and let each remaining original terminal absorb $\alpha_0/2$ units. A saturated flow certifies $\alpha_0/4$ well-linkedness. Otherwise its minimum cut identifies vertices to delete, and the decrease in the \emph{original} boundary pays for the lost terminals.

\begin{figure}[t]
\centering
\begin{tikzpicture}[x=1cm,y=1cm,font=\small,>=Latex]
  \node[draw,rounded corners,fill=blue!5,minimum width=3.8cm,minimum height=1.65cm] (host) at (6,0) {};
  \node[above=2pt] at (host.north) {current induced graph $C_0[U]$};
  \node[circle,draw,inner sep=4pt] (s) at (0.7,0) {$s$};
  \node[circle,draw,inner sep=4pt] (v) at (4.8,0) {$v$};
  \node[circle,draw,inner sep=4pt] (z) at (7.2,0) {$z$};
  \node[circle,draw,inner sep=4pt] (t) at (11.4,0) {$t$};
  \draw[->,thick] (s) -- node[pos=.38,above=6pt,align=center,text width=3cm] {original boundary\\supply $b_v$} (v);
  \draw[<->,thick] (v) -- node[above=5pt] {capacity $1$} (z);
  \draw[->,thick] (z) -- node[pos=.62,above=6pt,align=center,text width=3cm] {$\alpha_0/2$ at each\\surviving terminal} (t);
  \node[align=center,text width=12cm] at (6,-1.55) {A saturated flow certifies well-linkedness. A failed flow gives a deletion paid for by the decrease in the boundary of $U$ in $C_0$.};
\end{tikzpicture}
\caption{The trimming network. The drawing represents the types of arcs; the actual network has a supply arc at every boundary vertex and a sink arc at every surviving original terminal. The reference graph $C_0$ remains fixed.}
\label{fig:flow}
\end{figure}
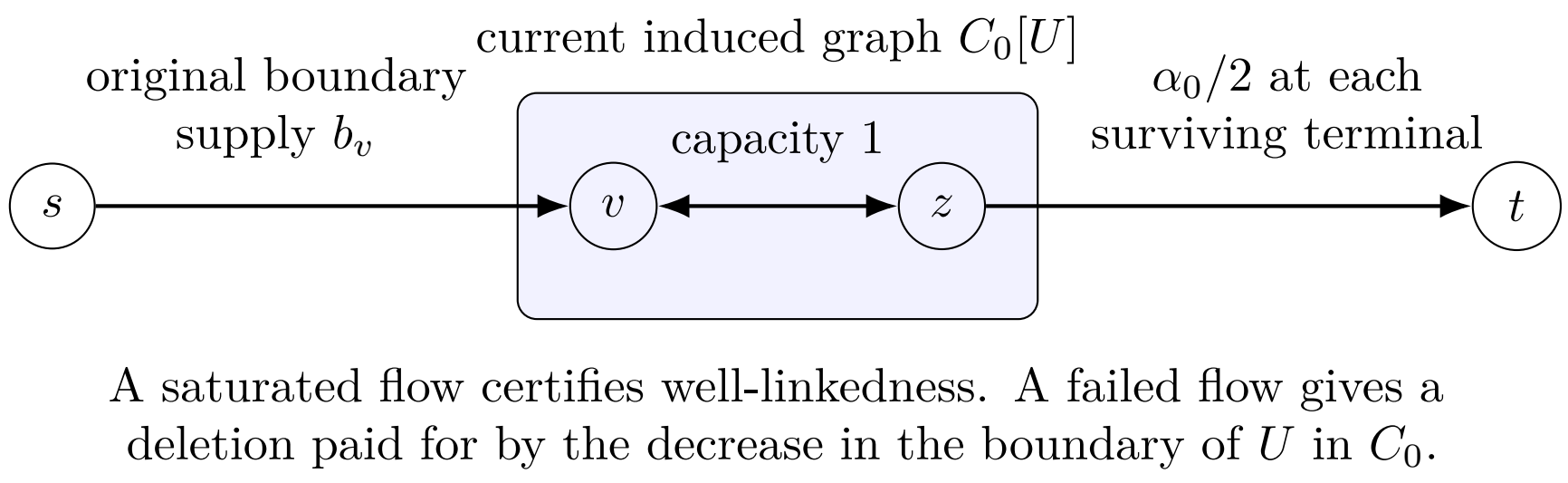

This accounting is over the entire trajectory of a cluster. Frontier restrictions increase its original boundary; trimming decreases it. All trimming losses are therefore charged to the sum of the small frontier separators. The connectivity guarantee stays $\alpha_0/4$ at every depth. After logarithmically many levels all clusters have been captured, and one depth carries enough total mass.

A captured port only touches its row. Routing needs the cluster to contain the \emph{whole} row interval in its cell. We expand all selected intervals simultaneously by decomposing their endpoint terminals at sparse cuts. Drop an interval's designation whenever any of its edges crosses a cut, retaining the induced hosts. Disjoint intervals charge distinct cut edges. Full well-linkedness also prevents an original port set from being dispersed among many final pieces without paying for many cuts. Thus a constant fraction of the total interval mass survives in disjoint, sufficiently large pieces.

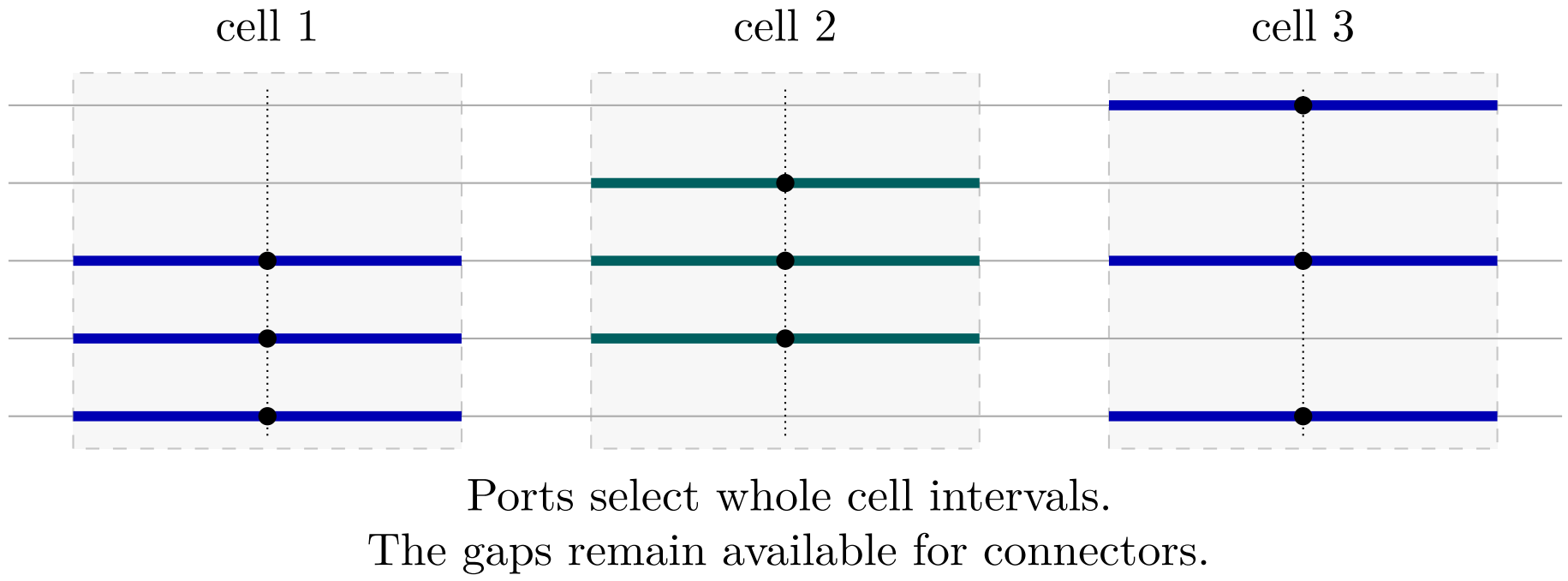
\begin{figure}[t]
\centering
\begin{tikzpicture}[x=1cm,y=1cm,font=\small]
  \foreach \a/\b/\lab in {0.5/3.5/1,4.5/7.5/2,8.5/11.5/3}{
    \fill[gray!6] (\a,-0.25) rectangle (\b,2.65);
    \draw[gray!45,dashed] (\a,-0.25) rectangle (\b,2.65);
    \node at ({(\a+\b)/2},3.02) {cell $\lab$};
  }
  \foreach \y in {0,0.6,1.2,1.8,2.4}{\draw[gray!65] (0,\y)--(12,\y);}
  \draw[blue!70!black,line width=2.2pt] (0.5,0)--(3.5,0);
  \draw[blue!70!black,line width=2.2pt] (0.5,0.6)--(3.5,0.6);
  \draw[blue!70!black,line width=2.2pt] (0.5,1.2)--(3.5,1.2);
  \draw[teal!75!black,line width=2.2pt] (4.5,0.6)--(7.5,0.6);
  \draw[teal!75!black,line width=2.2pt] (4.5,1.2)--(7.5,1.2);
  \draw[teal!75!black,line width=2.2pt] (4.5,1.8)--(7.5,1.8);
  \draw[blue!70!black,line width=2.2pt] (8.5,0)--(11.5,0);
  \draw[blue!70!black,line width=2.2pt] (8.5,1.2)--(11.5,1.2);
  \draw[blue!70!black,line width=2.2pt] (8.5,2.4)--(11.5,2.4);
  \foreach \x in {2,6,10}{\draw[densely dotted,black] (\x,-0.15)--(\x,2.55);}
  \foreach \x/\y in {2/0,2/0.6,2/1.2,6/0.6,6/1.2,6/1.8,10/0,10/1.2,10/2.4}{\fill (\x,\y) circle (2pt);}
  \node[align=center,text width=12cm] at (6,-0.85) {Ports select whole cell intervals.\\The gaps remain available for connectors.};
\end{tikzpicture}
\caption{An ordered family of cells from one recursion depth. Dotted lines indicate frontiers and dots indicate exported ports. Thick segments show selected whole row intervals. Clusters and their internal edges are suppressed; a final cluster contains every interval assigned to it.}
\label{fig:cells}
\end{figure}

Join successive assigned clusters along each row in an auxiliary multigraph. Its edges are realized by intervening row segments. Well-linkedness transfers large auxiliary cuts to large physical cuts. A counting argument then forces a chain of $\ell$ well-connected index intervals: otherwise longest-path ranks partition all assignments into fewer than $\ell$ classes, each accounting for only $O(N)$ assignments, where $N$ is the number of rows. Whole-interval containment makes consecutive routing regions overlap only in their common endpoint cluster. Maximum flow, followed by constructive boosting, yields the strong system. The required strong width is $H/\polylog m$ and the available mass is $m/\polylog m$, so $H=\widetilde O(w)$ and $m=\widetilde O(w\rho\ell)$ suffice.

\subsection{Constructing the smaller attachments}

Start from the constructive strong-chain supplier. In an odd cluster, add a temporary hub adjacent to every interface terminal and construct a routing hierarchy. Its connected pieces have well-linked full boundaries, and its parent routings connect child-boundary distributions with logarithmic congestion.

We seek a piece with boundary near the fourth-power scale that can send its boundary supply to the hub. It becomes the hair. Trim its complement against the original interface terminals, preserving a large connected base. Compute one unit-vertex-capacity maximum flow from the side's boundary to the surviving base. Accessibility and the fact that most original terminals survive give its cut inequalities directly. Truncate each path at its last departure from the side and first arrival in the base. Constructive boosting retains a proportional subset of \emph{each} terminal class, even when their sizes differ. The two backbone interfaces therefore remain large while the hair has the smaller required width.

If the hierarchy supplies no such piece, we construct a grid in its quotients. Away from the hub, a large cluster with small children gives an expanding quotient directly. The remaining case is a chain of large-boundary clusters ending at the hub, with small side pieces. Contract these pieces and consider three kinds of quotient: one carrying the original terminals, one at the largest boundary on the chain, and one for the children at each chain level. Delete the hub-containing piece before seeking a grid.

A common weighted routine returns a grid or a balanced separator. If every call returned a separator, the original-terminal quotient would bound the terminal mass, the largest-boundary quotient would bound the intermediate boundary, and the level quotients would bound the total side-piece mass. These bounds contradict the chosen backbone width. The routine itself grows connected grid branch sets by breadth-first search; when an existing branch set loses all access to unused vertices, it is discarded. Discarded vertices never return, giving a polynomial iteration bound.

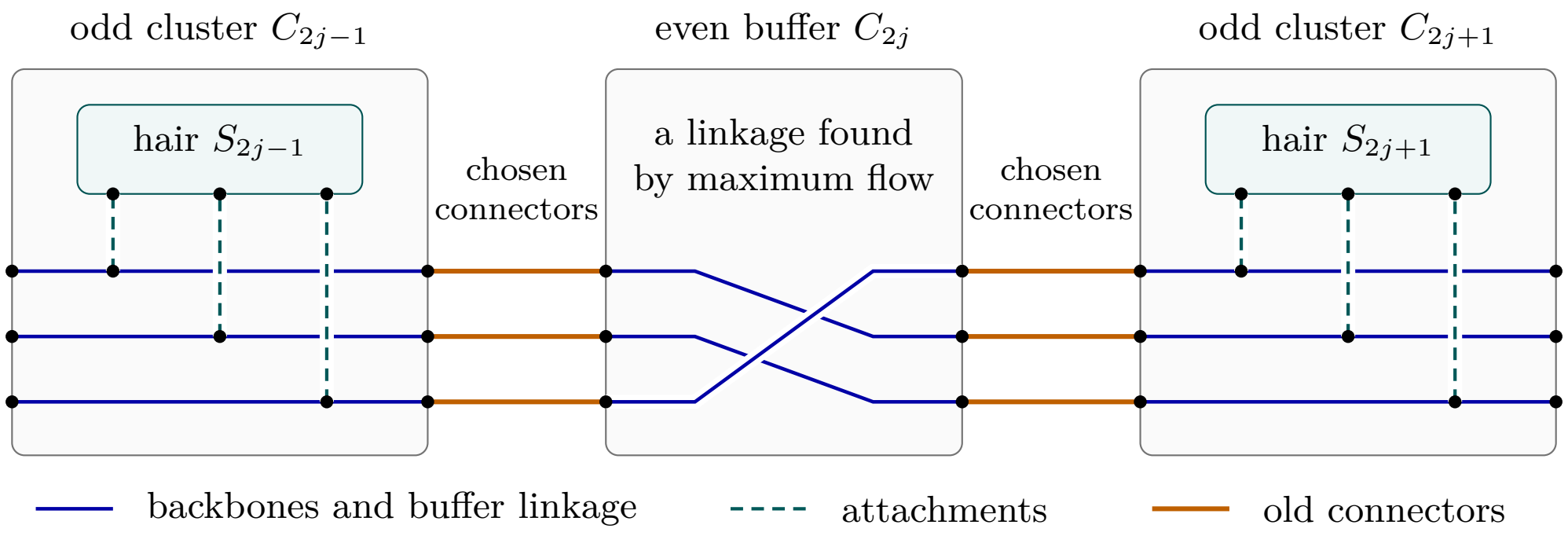
\begin{figure}[htbp]
\centering
\begin{tikzpicture}[
  x=1cm,y=1cm,font=\footnotesize,
  oldcluster/.style={draw=black!55,rounded corners=3pt,fill=black!2},
  hair/.style={draw=teal!65!black,rounded corners=3pt,fill=teal!6},
  backbone/.style={draw=blue!65!black,line width=.85pt},
  connector/.style={draw=orange!75!black,line width=1.15pt},
  attachment/.style={draw=teal!70!black,densely dashed,line width=.8pt,
    preaction={draw=white,line width=3.4pt}},
  port/.style={circle,fill=black,inner sep=0pt,minimum size=3.1pt},
  overpass/.style={preaction={draw=white,line width=3.4pt}}
]
\path[oldcluster] (0,-1.00) rectangle (3.5,2.25);
\path[oldcluster] (5.0,-1.00) rectangle (8.0,2.25);
\path[oldcluster] (9.5,-1.00) rectangle (13.0,2.25);
\node at (1.75,2.57) {odd cluster $C_{2j-1}$};
\node at (6.50,2.57) {even buffer $C_{2j}$};
\node at (11.25,2.57) {odd cluster $C_{2j+1}$};

\path[hair] (.55,1.20) rectangle (2.95,1.95);
\path[hair] (10.05,1.20) rectangle (12.45,1.95);
\node at (1.75,1.63) {hair $S_{2j-1}$};
\node at (11.25,1.63) {hair $S_{2j+1}$};
\node[align=center] at (6.50,1.50)
  {a linkage found\\by maximum flow};

\foreach \yy in {.55,0,-.55}{
  \draw[backbone] (0,\yy)--(3.5,\yy);
  \draw[backbone] (9.5,\yy)--(13.0,\yy);
  \draw[connector] (3.5,\yy)--(5.0,\yy);
  \draw[connector] (8.0,\yy)--(9.5,\yy);
}
\draw[backbone] (5.0,.55)--(5.75,.55)--(7.25,0)--(8.0,0);
\draw[backbone] (5.0,0)--(5.75,0)--(7.25,-.55)--(8.0,-.55);
\draw[backbone,overpass]
  (5.0,-.55)--(5.75,-.55)--(7.25,.55)--(8.0,.55);

\foreach \xx/\yy in {.85/.55,1.75/0,2.65/-.55,
                         10.35/.55,11.25/0,12.15/-.55}{
  \draw[attachment] (\xx,\yy)--(\xx,1.20);
  \node[port] at (\xx,\yy) {};
  \node[port] at (\xx,1.20) {};
}
\foreach \xx in {0,3.5,5.0,8.0,9.5,13.0}{
  \foreach \yy in {.55,0,-.55}{\node[port] at (\xx,\yy) {};}
}
\node[font=\scriptsize,align=center] at (4.25,1.25) {chosen\\connectors};
\node[font=\scriptsize,align=center] at (8.75,1.25) {chosen\\connectors};

\draw[backbone] (.20,-1.45)--(.85,-1.45);
\node[anchor=west] at (1.00,-1.45) {backbones and buffer linkage};
\draw[attachment] (6.05,-1.45)--(6.70,-1.45);
\node[anchor=west] at (6.85,-1.45) {attachments};
\draw[connector] (9.60,-1.45)--(10.25,-1.45);
\node[anchor=west] at (10.40,-1.45) {old connectors};
\end{tikzpicture}
\caption{Direct assembly inside the original chain. Three of the backbone paths are shown in each odd cluster. Keep only the old connectors incident with the selected backbone endpoints, and join their opposite endpoints by a linkage in the intervening even cluster. The pairing is chosen by the flow. Each hair remains inside its original odd cluster and meets its crossbar only at the attachment endpoints. White breaks indicate disjoint paths crossing in the drawing.}
\label{fig:direct-assembly}
\end{figure}

The original strong chain already supplies the global routing; see Figure~\ref{fig:direct-assembly}. Keep only the connectors incident with the selected crossbar endpoints and stitch them through the unchanged even clusters. The hairs realize the matchings requested by an explicit random-sign cut player. After $O(\log^2q)$ rounds, an exact matrix-potential test certifies expansion with constant probability. The weighted routine extracts a grid of side $\Omega(q/\log^{5/2}q)$. A degree-three treewidth sparsifier makes the working order polynomial in the promised threshold, so every logarithmic loss is a power of $\log q$. Choosing $q=g\,\polylog g$ completes the sixth-power construction.

\subsection{How a covering bound would lower the exponent}

The local input also admits an interpretation independent of row geometry. Index the two linkages by their common starting terminals in $A$. For each $a\in A$, the union of its $A$--$B$ path and its $A$--$X$ path contains a colorful tree, one meeting all three terminal sets. Any vertex belongs to at most two of these trees. Equivalently, assigning weight $1/2$ to each tree gives a fractional packing of total weight $\kappa/2$, with weight at most one through each vertex. Every vertex set meeting all colorful trees therefore has size at least $\kappa/2$.

Suppose a graph excluding the $g\times g$ grid satisfied a bound $\tau\le h(g)\nu$, up to logarithmic factors, between colorful covering and packing. The large transversal would then force about $\kappa/h(g)$ disjoint colorful trees. Each such tree supplies one crossbar component: retain its $A$--$B$ path and the branch from the chosen vertex of $X$ to that path. The leaf condition on $X$ makes the branch nontrivial. Different trees give disjoint components, and their terminal pairing can be arbitrary.

The assembly uses quadratically many backbones: a crossbar of order $q^2$ is needed in each of logarithmically many local instances, with $q=g\,\polylog g$. Thus it suffices to supply
\[
 \kappa=\widetilde O\bigl(g^2h(g)\bigr)
\]
terminals in each instance. This explains the entire exponent transfer. The factor $g^2$ comes from the number of backbones in the assembly; the covering factor measures the additional cost of obtaining a matching through each hair. The hairy-system supplier introduces only logarithmic losses at this length. A bound $h(g)=g^a\polylog g$ would therefore give grid exponent $a+2$. Every saving in the covering exponent produces the same saving in the final threshold.

The missing step is a three-terminal packing theorem. Menger's theorem controls paths between two terminal sets, whereas a colorful tree must meet all three. The two linkages already provide the fractional certificate above, even when their paths intersect in an arbitrary order. What is needed is a quantitative bound on the integral packing--covering loss in a graph excluding the target grid. Section~\ref{sec:conditional-cover} states this hypothesis for a general fixed exponent $a$. The case of linear dependence on $g$ gives the cubic exponent, up to logarithmic factors; no intermediate exponent needs a separate argument.

There is a separate algorithmic requirement. A covering inequality proves that a packing exists. A polynomial-time construction at an improved exponent would also need to find that packing and supply its linkage instances within the same threshold. In the unconditional sixth-power theorem, the local dichotomy gives an explicit alternative: it finds the crossbar, or converts the failed search into a strong path-of-sets system large enough to contain the grid.

\section{Preliminaries and external ingredients}
\label{sec:prelim}

All graphs are finite and undirected. Parallel edges created by contractions are retained when counting cuts; loops are discarded. Paths and linkages are vertex-disjoint unless another convention is stated. A linkage from $A$ to $B$ uses each vertex of both sets exactly once and does not prescribe their pairing. If the endpoint sets overlap, every common endpoint is a forced trivial path. A minor model is a family of disjoint connected branch sets with an edge between each pair required by the target graph. A model in a minor of $G$ lifts to a model in $G$.

We use base-two logarithms and write $\tO$ to suppress a fixed power of $\log(g+2)$, or of the displayed parameter product. All constants are absolute unless otherwise specified.

\paragraph{Full terminal well-linkedness.}
A terminal multiset $\Gamma$ is \emph{fully $\alpha$ edge-well-linked} in $C$ if every $U\subseteq V(C)$ satisfies
\begin{equation}\label{eq:full-wl}
 |E_C(U,V(C)\setminus U)|\ge
 \alpha\min\{\Gamma(U),\Gamma(V(C)\setminus U)\}.
\end{equation}
Here $\Gamma(U)$ counts occurrences. The word ``fully'' emphasizes that there is no cap on the terminal mass in the minimum. Every submultiset inherits the same inequality. A set is \emph{node-well-linked} if every two disjoint equal-sized subsets can be joined by a linkage. Two disjoint sets are \emph{linked} if all their equal-sized subsets can be joined in this way.

\paragraph{Rows and cells.}
A \emph{perfect unique row linkage} is a linkage whose paths cover the host graph and are the unique unpaired linkage between their first and last endpoints. Orient every row from first to last. A \emph{cell} contains one possibly empty interval of each row. A family of cells is ordered if its intervals are pairwise disjoint and appear in a common order on every row. When a cluster is assigned a row in its cell, it must contain that entire cell interval. Put one terminal occurrence at each end of an assigned interval. A singleton interval therefore contributes two occurrences at one vertex. This convention is needed when the same vertex is used by two consecutive auxiliary edge realizations.

\paragraph{Path-of-sets systems.}
A weak system of width $u$ and length $\ell$ consists of disjoint connected clusters $C_1,\ldots,C_\ell$, disjoint sets $A_i,B_i\subseteq V(C_i)$ of size $u$, and linkages from $B_i$ to $A_{i+1}$. All connector paths are mutually disjoint and their interiors avoid every cluster. We will specify a full edge-well-linkedness parameter for each $A_i\cup B_i$. A system is \emph{strong} if each $A_i$ and $B_i$ is node-well-linked in $C_i$ and $(A_i,B_i)$ is linked there.

A \emph{hairy} system additionally has disjoint hair clusters $S_i$ outside the backbone clusters and connectors, a node-well-linked set $Y_i\subseteq V(S_i)$, and a linkage $Q_i$ from a set $X_i\subseteq V(C_i)$ to $Y_i$. The sets $A_i,B_i,X_i$ are disjoint, $A_i$ and $X_i$ are linked, and all paths in the families $Q_i$ are mutually vertex-disjoint and disjoint from every backbone connector. Their interiors avoid every $C_j$ and $S_j$, and their endpoints belong only to their designated clusters. The standard hairy system has a common interface width. Our construction also uses unequal base and hair widths, with the same disjointness conditions; these widths are always specified.

\paragraph{Constructive suppliers.}
We use the following algorithmic results.
\begin{enumerate}[label=(\roman*)]
\item Given a graph of treewidth at least $F$, the degree-three sparsifier of Chekuri and Chuzhoy~\cite[Theorem~1.1]{CC15Sparsifiers} constructs, with high probability and in time polynomial in the input size and $F$, a topological minor $H$ with
\[
 |V(H)|\le C_sF^4,\qquad \Delta(H)\le3,\qquad
 \tw(H)\ge \frac{F}{C_s\log^{a_s}(F+2)},
\]
for absolute constants $C_s,a_s$.
\item The constructive path-of-sets theorem~\cite[Theorem~3.5]{CC16} supplies a strong system of width $w$ and length $L_0$ when the treewidth is at least $C_p wL_0^{48}\log^{a_p}(n+2)$, for suitable absolute constants $C_p,a_p$, where $n$ is the order of the input graph. Its running time is polynomial and it succeeds with high probability.
\item A strong path-of-sets system of width at least $16g^2+10g$ and length at least $2g(g-1)$ yields a $g\times g$ grid minor by a polynomial-time construction~\cite[Corollary~3.4]{CC16}.
\end{enumerate}
Increasing the fixed lower cutoff of the parameter makes each of the first two success probabilities at least $9/10$. Their published polynomial running-time guarantees give fixed clocks on arbitrary inputs. A trial may return failure when a promise is false; every grid exit is checked in the original graph.

\paragraph{Deterministic logarithmic approximation.}
In the deterministic part of the construction we use the cut algorithm of Linial, London, and Rabinovich~\cite[Theorem~4.1]{LLR95}, with a sufficiently large monotone bound $\varrho(k)=c\lceil\log_2(2k+2)\rceil$. Product demands on $k$ terminal occurrences have at most $k^2$ nonzero commodities; the logarithm of this number and the factor-two conversion to minority-terminal sparsity are absorbed in $c$. In the constructive proof of well-linkedness boosting~\cite[Appendix~A.1.2]{CC16}, replace the terminal sparsest-cut call in Theorem~A.3 by this deterministic routine. The argument uses the approximation ratio only through its stated guarantee; its remaining operations are grouping, spanning-tree computations, and integral flows. Thus the algorithmic losses become $32\Delta^4\varrho(k)$ for a single class and $1024\Delta^5\varrho(k)$ for partitioned boosting. The routing hierarchy is implemented with the same deterministic cut routine. This choice introduces no additional random failure events; it changes only the fixed logarithmic losses already allowed in our parameters.

\paragraph{Structural tools used only for the conditional reduction.}
The conditional results also use the hairy-system supplier and crossbar-chain theorem of Chuzhoy and Tan~\cite[Theorems~2.3 and~3.2]{CT21}. Treewidth $k$ supplies a subcubic hairy system of width $w$ and length $L$ when $k/\log^{c_h}k>C_hwL^{48}$. For a power of two $q$, crossbars of width $q^2$ in the odd clusters of a system of length $C_b\log q$ force a grid of side $c_bq/\log^{3/2}q$. These are used existentially in Section~\ref{sec:conditional-cover}; the unconditional algorithm below uses the constructive suppliers above and its own splitting procedure.

For a strong path-of-sets cluster with disjoint interfaces $A,B$ of size $k$, the union $T=A\cup B$ is fully $1/2$ edge-well-linked. Indeed, for a cut $(U,W)$ with $|T\cap U|\le |T\cap W|$, put $a=|A\cap U|$ and $b=|B\cap U|$. Then $a+b\le k$, so $|B\cap W|\ge a$ and $|A\cap W|\ge b$. Linkedness gives $a$ disjoint paths crossing the cut and, in a separate application, $b$ such paths. Hence $|E(U,W)|\ge\max\{a,b\}\ge|T\cap U|/2$. Here linkedness means linkage for every choice of equal-sized subsets, as in~\cite[Definition~3.1]{CC16}; we do not assume that $A\cup B$ is node-well-linked.

For the cut oracle, combine the two ordered demands of each unordered terminal pair. This leaves the concurrent-flow value and every cut's total demand unchanged, and puts the instance in the standard undirected-demand form of~\cite[Theorem~4.1]{LLR95}.

\subsection{Polynomial terminal-cut search}
\label{sec:cut-oracle}

We need approximate terminal expansion, with a ratio depending on terminal mass rather than host size. Let $K$ bound that mass. Fix a nondecreasing function
\[
 r(K)=c_{\rm sc}\lceil\log_2(K+2)\rceil\ge1
\]
large enough for the following deterministic algorithmic guarantee. Given a graph and terminal multiset $\Gamma$ of mass at most $K$, a polynomial-time routine returns a cut with terminal expansion at most $r(K)$ times the minimum finite terminal expansion. If all occurrences lie at one vertex, the well-linkedness condition is vacuous and no cut is needed.

This guarantee follows from the multicommodity sparsest-cut algorithm of Linial, London, and Rabinovich~\cite[Theorem~4.1]{LLR95}. To see the relevant dependence, put demand $\mu_u\mu_v/|\Gamma|$ between each pair of distinct terminal locations, where $\mu_v$ is its multiplicity. There are at most $K^2$ demand pairs. A cut with terminal masses $x$ and $|\Gamma|-x$ has crossing demand $x(|\Gamma|-x)/|\Gamma|$, which lies between one half and one times the smaller terminal mass. The $O(\log K)$ approximation therefore applies also to~\eqref{eq:full-wl}. We enlarge $r$ to cover both this routine and the terminal-count approximation factors used in algorithmic boosting.

\begin{lemma}[Decomposition with a cut budget]\label{lem:decomp}
Start with a terminal multiset of mass $K\ge2$. Recursively split a piece only at a cut whose size is less than $\eta$ times its smaller current terminal mass. Terminal occurrences may be discarded but never introduced. The total number $b$ of cut edges satisfies
\[
 b\le\eta K\log_2 K.
\]
If the routine above is used and splitting stops when its returned ratio is at least $\eta$, every final terminal set is fully $\eta/r(K)$ edge-well-linked. The decomposition is polynomial-time computable.
\end{lemma}
\begin{proof}
At a split, the smaller current terminal mass is at most the smaller mass of the original occurrences in the two children, including occurrences that have since lost their designation. Charge the latter quantity to those original occurrences. Each charged occurrence moves to a child with at most half its parent's original mass, so receives at most $\log_2K$ charges. This proves the cut budget. At termination, an $r(K)$-approximate ratio of at least $\eta$ implies that every ratio is at least $\eta/r(K)$. Each split increases the number of nonempty vertex parts, so at most $|V|-1$ splits occur. Each cut search takes polynomial time.
\end{proof}

\section{From two linkages to well-linked clusters}
\label{sec:source}

\subsection{Minimum-cost preprocessing}

\begin{lemma}[Minimum-cost linkage relative to a fixed subgraph]
\label{lem:mincost-linkage}
Let $F$ be a fixed subgraph of an undirected graph $H$, let $S,T$ be disjoint vertex sets, and suppose an $r$-linkage from $S$ to $T$ exists. Give each edge of $F$ cost zero and every other edge cost one. One minimum-cost integral vertex-flow computation constructs an $r$-linkage $\mathcal L$ such that, in $J=F\cup\mathcal L$, every edge of $E(\mathcal L)\setminus E(F)$ belongs to every $r$-linkage from $S$ to $T$. In particular, these edges are essential when the endpoint sets selected by $\mathcal L$ are fixed.
\end{lemma}
\begin{proof}
Use a directed vertex-split network with capacity one at every vertex, including terminals. Replace each undirected edge by its two directed copies, both with capacity one and the stated cost. Add capacity-one, cost-zero arcs from a new source to $S$ and from $T$ to a new sink. Retain ordinary graph arcs at terminal vertices, so an unused terminal may be internal to a flow path. Compute a minimum-cost integral flow of exactly value $r$ and delete its directed cycles. Nonnegative costs ensure that cycle deletion preserves optimality. The resulting simple, vertex-disjoint paths have cost $|E(\mathcal L)\setminus E(F)|$.

Every competing $r$-linkage in $J$ uses only edges of $F$ and edges of $E(\mathcal L)\setminus E(F)$. Omitting any edge of the latter set would strictly lower the cost, contradicting optimality in $H$. This argument also applies when the optimization permits the endpoint subsets to vary.
\end{proof}

After cycle deletion, vertex-disjoint simple paths cannot use both orientations of an undirected edge. Thus no shared edge-capacity gadget is needed. Parallel-edge identities are retained and loops are discarded. Successive shortest augmentations, with Bellman--Ford in the residual network, implement this computation in $O(rn(n+e))$ time on an $n$-vertex, $e$-edge input; there are at most $r\le n$ augmentations.

All linkage problems in this preprocessing are unpaired, so a single-commodity minimum-cost flow suffices.

\begin{lemma}[An essential-row core]\label{lem:essential-core}
Let $F$ be a graph whose components are paths, possibly trivial, and let
$\mathcal L$ be a full unpaired $(S,T)$-linkage, where $S,T$ are disjoint
and $S\cap V(F)=\varnothing$. Put $J=F\cup\bigcup_{P\in\mathcal L}P$.
Suppose every edge of $E(\mathcal L)\setminus E(F)$ belongs to every
full $(S,T)$-linkage in $J$.
In polynomial time one can construct a degree-four minor $J^\star$
covered by the images $\mathcal L^\star$, which form the unique
unpaired linkage between the images $S^\star,T^\star$.
These endpoint sets remain disjoint and retain their cardinalities.
The nonempty images of the components of $F$ are mutually disjoint paths,
and every incidence between an $F$-component and a row is preserved.
\end{lemma}
\begin{proof}
Keep parallel-edge identities and the two family memberships.
First contract every edge common to $F$ and $\mathcal L$.
Then repeatedly contract an $F$-edge with an endpoint outside the rows,
and finally discard components disjoint from the rows.
The first operation contracts common row segments; the second never
merges two row vertices. Hence the two families remain
vertex-disjoint path families, their incidences are preserved, and
their union has maximum degree four. At termination every vertex
lies on a row: any nonisolated off-row vertex would permit another
contraction, and an isolated one would be discarded.

Every contraction branch set meets at most one original row, in a
contiguous segment, and is connected using only edges of $F$.
Vertices of $S$ are untouched because they lie outside $F$.
Thus no row collapses to a singleton, no endpoints are identified,
and $S^\star,T^\star$ remain disjoint.

If a full $(S^\star,T^\star)$-linkage $\mathcal Z$ differs from
$\mathcal L^\star$, it omits a row edge. Indeed, using every row
edge already gives degree two at each internal row vertex and degree
one at each endpoint; since the rows cover the host, no additional
linkage edge is possible. Let $e$ be the original edge corresponding
to an omitted row edge. It lies outside $F$, because all common edges
were contracted. Lift $\mathcal Z$ inside the disjoint branch sets,
joining its incident edge endpoints, or its prescribed terminal,
using only edges of $F$. This gives a full $(S,T)$-linkage in $J$
avoiding $e$, contrary to essentiality.

There are at most $|V(J)|$ contractions. The path images, incidences,
and contraction branch sets can all be maintained in polynomial time.
\end{proof}

\subsection{Marking, pruning, and decomposition}

\begin{lemma}[Marked-block source]\label{lem:source}
Let $N,m,H\ge1$ be integers with $m\ge N$. Suppose $N$ disjoint rows cover a maximum-degree-four graph, and $m$ disjoint auxiliary paths each meet at least $64H$ distinct rows. There is a polynomial-time construction of disjoint connected clusters whose terminal sets are fully $\alpha_0$ edge-well-linked, have total mass $T_0\in[m,16m]$, and each have at least $8H$ terminals, where
\[
 \alpha_0=\frac{1}{1000\,r(16m)\lceil\log_2(2B)\rceil},
 \qquad 3m\le B\le8m.
\]
Here $B$ is an integer produced by the construction. The rows need not be unique for this lemma.
\end{lemma}
\begin{proof}
Choose $64H$ marked contacts on each auxiliary path, one on each of $64H$ distinct rows. On every row, group its marks into consecutive blocks of $8H$, discarding the final incomplete block. Realize each full block by the interval between its first and last marks. The intervals are disjoint, with distinct endpoints. Their number $B_0$ satisfies
\[
 8m-N\le B_0\le8m.
\]
Form the bipartite marked-incidence graph between blocks and auxiliary paths. Each block has degree $8H$, and each auxiliary path has degree at most $64H$. There is at most one marked incidence between a path and a block.

Prune blocks of degree below $4H$ and paths of degree below $16H$ until neither exists. If $e,b,t$ denote the current numbers of incidences, blocks, and paths, the potential
\[
 \Phi=e-4Hb-16Ht
\]
does not decrease. Initially it is $4HB_0-16Hm\ge16Hm-4HN$. Finally it is at most $4Hb$, since $e\le8Hb$. Thus at least $4m-N\ge3m$ blocks survive; call their number $B$.

Place a terminal at each end of every surviving block. Apply Lemma~\ref{lem:decomp} with threshold $\eta=1/(1000\lceil\log_2(2B)\rceil)$. Whenever a designated block crosses a cut, discard its designation and both endpoint occurrences. Keep the full induced host on each child, including the remaining portions of discarded objects. The total cut size is
\[
 b_{\rm cut}\le2\eta B\log_2(2B)\le B/500.
\]
At most $b_{\rm cut}$ blocks and at most $b_{\rm cut}$ auxiliary paths cross the cuts, because each family is disjoint. Remove these objects from the incidence graph. At most $72H b_{\rm cut}$ incidences disappear.

Before these removals, the potential $\Psi=e-Hb-4Ht$ is at least $e/2\ge2HB$: both subtracted terms are at most $e/4$. Now prune blocks of degree below $H$ and paths of degree below $4H$. This cannot decrease $\Psi$. If $B'$ blocks survive, then
\[
 7HB'\ge2HB-72Hb_{\rm cut}>7HB/4,
 \qquad B'\ge B/4.
\]
Their endpoint mass is therefore at least $2B'\ge B/2\ge m$ and at most $2B\le16m$.

Every final piece containing a retained block also contains a retained whole auxiliary neighbor. That path has at least $4H$ retained block neighbors, all in the same piece, so the piece has at least $8H$ retained terminals. They are a subset of its final decomposition multiset, and therefore inherit full $\alpha_0$ well-linkedness. Retain the connected component containing them. It contains all the retained blocks and auxiliary neighbors used in the count. The constructions and both degree-pruning passes are polynomial-time operations.
\end{proof}

\section{Frontiers in a perfect unique linkage}
\label{sec:frontier}

\begin{lemma}[Hereditary uniqueness]\label{lem:hereditary}
In a graph covered by a perfect unique row linkage, the components of the rows in any induced vertex set form a perfect unique unpaired linkage between their local first and last endpoints. Deleting non-row edges preserves this property.
\end{lemma}
\begin{proof}
Orient the rows. Splice a proposed alternative local linkage into their unchanged outside portions. The result consists of disjoint original-start--original-end paths and possibly directed cycles. The path components form an original unpaired linkage. By uniqueness they are the original spanning rows, leaving no vertex for a cycle or a changed local segment. Thus the local linkage is unchanged. Common local endpoints are forced trivial paths. Deleting edges cannot create an alternative linkage.
\end{proof}

\begin{lemma}[Balanced frontier]\label{lem:sweep}
If $N$ rows form a perfect unique linkage, they admit a sweep with one frontier vertex on each row, separating the vertices strictly before it from those strictly after it. Consecutive states advance one frontier by one row edge. For any $T$ distinct marks, some state has at most $\lceil T/2\rceil$ marks strictly on either side. The sweep is polynomial-time computable.
\end{lemma}
\begin{proof}
Start with the first vertex of every row. Put an arc $i\to j$ when an off-row edge joins the current frontier on row $i$ to a vertex strictly ahead of the frontier on row $j$. A directed cycle would permit cyclic exchanges of suffixes, yielding another unpaired linkage. A forward chord on one row similarly gives a shortcut. Thus the dependency digraph is acyclic and has an unfinished sink whenever an unfinished row remains; a finished row cannot be the target of an arc.

Advance such a sink. Its former frontier has no future neighbor except the next row vertex, so the separation invariant is preserved. Last vertices may stay on the frontier. The number of marks strictly behind it increases by at most one per step. Use the first state reaching $\lfloor T/2\rfloor$ such marks, if it exists; otherwise use the final state, whose future is empty. Both sides then satisfy the claimed bound. Scanning the edges after each advance gives a polynomial implementation.
\end{proof}

\section{Flow trimming against a fixed well-linked reference}
\label{sec:flow-trimming}

For a graph $C_0$ and $U\subseteq V(C_0)$, write
$\partial_EU=E_{C_0}(U,V(C_0)\setminus U)$ and
$b(U)=|\partial_EU|$.  The reference graph $C_0$ and its original
terminal set $\Gamma_0$ will remain fixed while $U$ shrinks.  All
well-linkedness statements in the next lemma use edge cuts.

\begin{lemma}[Trimming by one maximum flow]\label{lem:flow-trimming}
Suppose $\Gamma_0$ is fully $\alpha_0$ edge-well-linked in $C_0$,
where $0<\alpha_0\le1$.  For any $U\subseteq V(C_0)$, one
maximum-flow computation finds $U'\subseteq U$ such that
$\Gamma_0\cap U'$ is fully $\alpha_0/4$ edge-well-linked in
$C_0[U']$ and
\[
 b(U')\le b(U)-\frac{\alpha_0}{2}
                  |\Gamma_0\cap(U\setminus U')|.
\]
Moreover, adding to $\Gamma_0\cap U'$ all vertices of $U'$ incident
to $\partial_EU'$ gives a fully $\alpha_0/7$ edge-well-linked set
in $C_0[U']$.
The computed flow sends one unit from each boundary incidence to
$\Gamma_0\cap U'$, with unit edge congestion and sink capacity
$\alpha_0/2$ at each terminal.
If any terminals remain, $U'$ can be restricted to their connected
component.  This is the terminal-weight version of the exact-flow
trimming argument of Saranurak and Wang~\cite[Sections 3.2--3.3]{SW19}.
\end{lemma}
\begin{proof}
For $v\in U$, write $b_v=|E_{C_0}(\{v\},V(C_0)\setminus U)|$.
Construct a network with a source arc $sv$ of capacity $b_v$,
two opposite capacity-one arcs for each edge of $C_0[U]$, and a
sink arc $vt$ of capacity $\alpha_0/2$ at each terminal
$v\in\Gamma_0\cap U$.  Compute a maximum flow $f$, let
$\{s\}\cup A$ be its source-reachable residual set, and put
$U'=U\setminus A$.  The minimum-cut inequality gives
\[
 b(U)-\sum_{v\in A}b_v
 +|E_{C_0}(A,U')|+\frac{\alpha_0}{2}|\Gamma_0\cap A|
 \le b(U).
\]
Rearranging and using
$b(U')=b(U)-\sum_{v\in A}b_v+|E_{C_0}(A,U')|$
proves the boundary and terminal-loss bound.

Every source arc into $U'$ and every arc from $A$ to $U'$ is
saturated, while the opposite cut arcs carry no flow.  After cancelling opposite flows on internal undirected edges, restricting
$f$ to $U'$ gives a feasible flow whose sources have
capacities $|E_{C_0}(\{v\},V(C_0)\setminus U')|$ and whose
terminal sinks still have capacity $\alpha_0/2$, with every source
saturated.  For a cut $Z,U'\setminus Z$, choose the side $Z$
containing the smaller number $z$ of retained terminals.  If $e$
is its internal edge boundary and $b_Z$ its boundary toward
$V(C_0)\setminus U'$, flow conservation gives
$b_Z\le e+\alpha_0z/2$.  Because $z\le|\Gamma_0|/2$, the
original well-linkedness gives $e+b_Z\ge\alpha_0z$.
Consequently $e\ge\alpha_0z/4$.

If $p$ is the number of boundary vertices in $Z$, then
$p\le b_Z\le e+\alpha_0z/2\le3e$. The union of the retained
terminals and the boundary vertices therefore has at most
\[
 z+p\le(4/\alpha_0+3)e\le7e/\alpha_0
\]
vertices on this side. Its smaller side has no greater terminal
count, proving the additional well-linkedness guarantee.

Positive well-linkedness puts all retained terminals in one
connected component. A terminal-free component has no sink, so
the saturated-boundary flow gives it zero boundary supply.
Removing these components preserves the flow certificate and both
well-linkedness guarantees, loses no terminal, and cannot increase
the original boundary.
\end{proof}

\paragraph{The trimming algorithm.}
Build the network in the proof, compute a maximum flow, delete
the original vertices reachable from its source in the residual
network, and retain the component containing the remaining
terminals.  The reference graph $C_0$ and terminal set $\Gamma_0$
remain unchanged in every future call.  If no terminal remains,
return the empty set.  In localization, the terminal-loss budget
proves that this last case never occurs.

\paragraph{Why the reference must remain fixed.}
The guarantee after trimming is $\alpha_0/4$ regardless of how many
earlier trimming calls were made.  The proof always compares a cut
with the original set $\Gamma_0$ in $C_0$.  Replacing that reference
by the output of the preceding call would incorrectly introduce a
factor four at every recursion level.

\section{Constructive recursive localization}
\label{sec:constructive-localization}

The next theorem places a large fraction of the terminal mass on frontiers in one ordered family of cells. Section~\ref{sec:expansion} then expands the exported ports to whole row intervals.

\begin{theorem}[Localization by maximum flow]
\label{thm:flow-localization}
Let $W\ge1$ be an integer, and let a graph of maximum degree $\Delta\ge3$ be covered by $N$
perfect unique unpaired rows.  Suppose it contains disjoint
connected subgraphs $C_0$, each equipped with a fully
$\alpha_0$ edge-well-linked set $\Gamma_0$, for a common
$0<\alpha_0\le1$.  Put
\[
 T_0=\sum_{C_0}|\Gamma_0|,\qquad
 d=\lceil\log_2(T_0+1)\rceil+2,\qquad
 a=\frac{\alpha_0}{4},\qquad
 \delta=\frac{\alpha_0}{64\Delta d}.
\]
If every original terminal set has size at least $128W/\delta$,
then maximum-flow computations and frontier sweeps find ordered,
vertex-disjoint cells containing disjoint connected clusters with
port sets having the following properties:
\begin{enumerate}
\item Every port lies on the cell's frontier, and within one cell
      no two ports lie on the same original row.
\item Each port set has at least $32W$ vertices and is fully
      $\beta$ edge-well-linked in its cluster, where
      $\beta=\alpha_0/8$.
\item The total number of ports is at least $\delta T_0/(4d)$.
\end{enumerate}
The construction uses polynomially many maximum-flow computations.
\end{theorem}
\begin{proof}
Every original cluster keeps its own fixed reference
$(C_0,\Gamma_0)$.  Its current cluster is an induced connected
subgraph $C_0[U]$, its current terminals are $\Gamma_0\cap U$,
and its current well-linkedness parameter is always at least $a$.
Initially $U=V(C_0)$, so the condition holds.

Each recursive cell has the induced host on its row intervals; the frontier is excluded from both child cells. At a recursion node, sweep its row intervals to a frontier $S$
that balances the union of all current terminal sets.  For a
current cluster $C=C_0[U]$ with $M$ current terminals, compute a
maximum vertex-disjoint linkage from $\Gamma_0\cap U$ to $S\cap U$,
allowing trivial paths.  If its value is at least $\delta M$,
capture the cluster, retaining $\lfloor\delta M\rfloor$ paths and
their distinct endpoints on $S$.

Otherwise, Menger's theorem gives a separating vertex set $K$
with $|K|<\delta M$.  Set $s=\Delta|K|$.  There is a component
$D$ of $C-K$ containing more than $M/2$ current terminals.  Indeed,
if no component did, a union of components would contain between
$M/4$ and $M/2$ terminals: either one component already lies in
this interval, or add components of terminal size below $M/4$
until reaching it.  The terminals in $K$ cannot prevent this,
since $|K|<\delta M<M/4$.  The resulting cut has at most $s$
edges, whereas current $a$-well-linkedness requires at least
$aM/4$; this contradicts $\Delta\delta<a/4$.

The component $D$ misses at most $s/a=4s/\alpha_0$ current
terminals.  This follows by applying current well-linkedness to
the cut $D,V(C)\setminus D$.  Its boundary in the original graph
is at most $b(U)+s$.  Apply \textsc{FlowTrim} to $D$, still using
$C_0,\Gamma_0,\alpha_0$ as the reference.  This restores
$a$-well-linkedness, and does not increase that original boundary.
Since $K$ separates the terminals from $S$, the component $D$
avoids $S$ and lies wholly on one side of the frontier.  Send the
trimmed cluster to that child cell.  Current clusters remain
disjoint.

Here is the loss accounting along one original cluster's
trajectory.  Let $M_0=|\Gamma_0|$.  Over at most $d$ levels, let
$S_{\rm tot}$ be the sum of the quantities $s$ just defined.
Since $M\le M_0$,
\[
 S_{\rm tot}\le d\Delta\delta M_0.
\]
Frontier-component restrictions lose at most
$4S_{\rm tot}/\alpha_0$ terminals.  These restrictions increase
the original boundary by at most $S_{\rm tot}$ altogether.
Each subsequent flow deletion decreases that boundary by at
least $\alpha_0/2$ times its terminal loss.  The boundary starts
at zero and stays nonnegative; hence all flow deletions together
lose at most $2S_{\rm tot}/\alpha_0$ terminals.  In total, at every
prefix of the trajectory, the terminal loss is at most
\[
 \frac{6S_{\rm tot}}{\alpha_0}
 \le\frac{6d\Delta\delta}{\alpha_0}M_0
 =\frac{3M_0}{32}<\frac{M_0}{8}.
\]
In particular, the trimmed cluster is never empty and always
retains more than half its original terminals.  This bound also
justifies every preceding invocation during the first $d$ levels.

The frontier balances the current terminals, so each child node
has at most the ceiling of half its parent's terminal count.
At level $d$ that count would be at most one.  But every surviving
cluster has more than half its original terminals, hence at least
two.  Therefore the recursion stops before level $d$, with every
surviving cluster captured.  The captured terminal mass is at least
$29T_0/32>T_0/2$.  One depth captures at least $T_0/(2d)$ terminals.
Its node cells are disjoint and ordered on every original row.

Fix that depth.  A captured cluster with $M$ terminals supplies
$p=\lfloor\delta M\rfloor\ge\delta M/2\ge32W$ distinct ports.
The internal export paths preserve well-linkedness up to a
constant factor: for a cut crossed by $t$ edges, at most $t$
export paths cross the cut, and if the smaller port side has $m$
ports, each side contains at least $m-t$ source terminals.
Thus $t\ge a(m-t)$, giving port well-linkedness
$a/(1+a)\ge\alpha_0/8$.
The total port mass at the selected depth is at least
$\delta T_0/(4d)$.  A frontier has one vertex on each current row,
and the captured clusters and export paths are disjoint, proving
the distinct-row property.
\end{proof}

\paragraph{Operation count.}
Let $n,m$ denote the size of the ambient host and let $t$ be the
number of original clusters.  Each original cluster participates
in at most $d$ recursion levels.  At each level, it requires one
frontier-linkage computation and at most one call to
\textsc{FlowTrim}; the latter uses one maximum flow.  There are
therefore $O(td)$ maximum-flow calls, each on a network with
$O(n)$ vertices and $O(n+m)$ arcs.  If $\mathsf{MF}(n,m)$ bounds
the running time of the chosen maximum-flow routine, these calls
cost $O(td\,\mathsf{MF}(O(n),O(n+m)))$.
Frontier sweeps and bookkeeping are polynomial; a direct
implementation by scanning the current edges after every frontier
advance takes at most $O(dn(n+m))$ additional time.  All graphs
and flow networks, including the fixed reference boundary counts,
are explicitly represented.

\section{Simultaneous expansion to whole row intervals}
\label{sec:expansion}

The ports obtained above lie on different rows in each cell. We now make the corresponding entire cell intervals belong to disjoint clusters. The induced hosts must remain intact when a designation is dropped: deleting the other portions of an interval or auxiliary path would invalidate the cut accounting.

\begin{lemma}[Whole-interval expansion]\label{lem:expansion}
Let $W,R\ge1$ be integers. Let a host contain $R$ disjoint row intervals, each with one distinguished port. Partition the ports into sets $P_1,\ldots,P_t$ contained in disjoint connected subgraphs $C_1,\ldots,C_t$. Suppose every $P_i$ is fully $\beta$ edge-well-linked in $C_i$, where $0<\beta\le1$, and $|P_i|\ge32W$.

Let $r$ dominate the terminal-cut approximation for mass $2R$, and let
\[
 0<\gamma\le\frac{\beta}{100r\log_2(2R)}.
\]
There is a polynomial-time construction of disjoint connected clusters with at least $R/2$ assigned whole intervals in total, at least $4W$ assigned intervals per cluster, and fully $\gamma$ edge-well-linked endpoint multisets. Singleton intervals are allowed.
\end{lemma}
\begin{proof}
Designate all selected intervals and put two endpoint occurrences on each. Decompose their endpoints using Lemma~\ref{lem:decomp} with threshold $r\gamma$. When any edge of an interval crosses a splitting cut, drop its designation and both endpoint occurrences. Endpoint locations alone do not determine whether an interval crosses. Keep the induced hosts on the children, without deleting the remaining parts of discarded intervals. The total cut size satisfies
\[
 b\le2r\gamma R\log_2(2R)\le\beta R/50.
\]
At most $b$ intervals lose their designation, since disjoint crossing intervals use distinct cut edges. They account for at most $b$ distinguished ports.

Let $V_j$ be the final vertex parts. Put $M_i=|P_i|$, $p_{ij}=|P_i\cap V_j|$, and let $b_i$ count cut edges inside $C_i$. Summing its full well-linkedness inequality over the final parts gives
\[
 \sum_j\min\{p_{ij},M_i-p_{ij}\}\le2b_i/\beta.
\]
The $C_i$ are disjoint, so $\sum_i b_i\le b$. If a part contains more than half of $P_i$, call it the dominant part of that port set. Thus all ports outside their dominant part, together with all ports of sets having no dominant part, number at most $2b/\beta$.

Call a final part bad if it contains $r_j<4W$ still-designated intervals. If it is dominant for some $P_i$, it contains more than $16W$ ports of that set. At most $r_j$ belong to retained intervals, so more than $12W$ belong to lost intervals. Its $r_j$ retained intervals can therefore be charged to lost-interval ports in that part at rate at most one third. Summed over these bad parts, this costs at most $b/3$ retained intervals. In every other bad part all retained ports are nondominant, costing at most $2b/\beta$ retained intervals.

The good parts consequently retain at least
\[
 R-b-b/3-2b/\beta\ge R/2
\]
designated intervals. Their endpoint multisets are fully $\gamma$ well-linked. In each good part retain the connected component containing those endpoints. Every assigned interval lies in that component. The decomposition and the final accounting are polynomial-time computable.
\end{proof}

\begin{corollary}[Ordered whole-interval mass]\label{cor:ordered-mass}
Under the hypotheses of Theorem~\ref{thm:flow-localization}, put
\[
 \beta=\alpha_0/8,
 \qquad
 \gamma=\frac{\beta}{100r(16m)\lceil\log_2(2N+2)\rceil},
\]
where $m\ge N$ and $T_0\le16m$. One can construct, in polynomial time, an ordered family of cell-supported disjoint clusters, each with at least $4W$ assigned whole cell-row intervals and full $\gamma$ endpoint well-linkedness, with total assignment mass at least $\delta T_0/(8d)$.
\end{corollary}
\begin{proof}
Use the depth selected in Theorem~\ref{thm:flow-localization}. Within each cell the exported ports lie on distinct rows. Select the whole interval on every such row and apply Lemma~\ref{lem:expansion}. There are at most $N$ selected intervals in a cell, so the common $\gamma$ is admissible. Each cell retains at least half its ports, and the cells are disjoint and ordered.
\end{proof}

\section{Ordered routing and constructive strongification}
\label{sec:ordered}

\begin{lemma}[Ordered routing]\label{lem:ordered}
Let $W\ge1$ and $\Delta\ge3$ be integers. Let a maximum-degree-$\Delta$ host contain disjoint clusters supported in ordered cells on $N$ rows. Every cluster contains its whole cell interval on every assigned row, has at least $4W$ assigned rows, and has a fully $\gamma$ edge-well-linked endpoint multiset, with $0<\gamma\le1$. If the total assignment mass exceeds $2N(\ell-1)$, a polynomial-time algorithm finds a weak path-of-sets system of length $\ell$ and width
\[
 u=\left\lfloor\frac{\gamma W}{3\Delta}\right\rfloor,
\]
with disjoint incoming and outgoing terminal sets whose unions are fully $\gamma$ edge-well-linked. The statement is nontrivial when $u\ge1$.
\end{lemma}
\begin{proof}
Order the clusters by cells, arbitrarily within one cell, and denote them $C_1,\ldots,C_s$. Form a row-labelled multigraph $F$ by joining consecutive clusters assigned each row. Realize each auxiliary edge by the row segment from the last endpoint of the earlier assigned interval to the first endpoint of the later one. These segments are edge-disjoint. Two consecutive segments can share an endpoint when an assigned interval is a singleton; its two terminal occurrences pay for its two auxiliary incidences.

For $a\le b$, let $J[a,b]$ be the exact edge union of clusters $C_a,\ldots,C_b$ and the realizations of the edges of $F[a,b]$. Shared physical edge identifiers are included once. Whole-interval containment has three consequences:
\begin{enumerate}[label=(\roman*)]
\item $J[a,b]$ avoids every cluster outside $[a,b]$;
\item a connector can leave $C_a$ only at an assigned last endpoint and enter $C_b$ only at an assigned first endpoint;
\item $J[a,b]\cap J[b,c]\subseteq C_b$ when $a<b<c$.
\end{enumerate}
To check the only subtle case, a segment with an endpoint in an endpoint cell uses a row assigned to that endpoint cluster. That cluster contains the whole cell interval on the row, excluding all other clusters in the same cell from it. Thus arbitrary tie-breaking within a cell causes no intersection problem.

\paragraph{From auxiliary connectivity to physical connectivity.}
Suppose $F[a,b]$ has $a$--$b$ edge-connectivity at least $2W$. Consider a physical cut of size $z$ separating all assigned last endpoints of $C_a$ from all assigned first endpoints of $C_b$. Orient every auxiliary cluster toward a side containing at least half of its endpoint mass. The endpoint clusters can be oriented to the prescribed source and target sides: each contributes one last and one first occurrence per assigned row. For the other clusters, full well-linkedness bounds the total minority mass by $z/\gamma$, since their internal cut edges are disjoint.

Each crossing auxiliary edge either crosses the physical cut on its realizing segment, or has an endpoint occurrence on its cluster's minority side. The first class has size at most $z$ and the second at most $z/\gamma$. Hence $z<\gamma W$ would give an auxiliary cut smaller than
\[
 z+z/\gamma\le2z/\gamma<2W,
\]
a contradiction. Physical edge-connectivity is therefore at least $\gamma W$. A vertex separator of size $k$ yields an edge cut of at most $\Delta k$ edges, so vertex Menger gives at least $\lfloor\gamma W/\Delta\rfloor$ disjoint paths between the appropriate endpoint sets. Trim each path at its last departure from $C_a$ and first subsequent arrival in $C_b$. By (ii), the endpoints remain assigned last and first endpoints, respectively.

\paragraph{The mass argument.}
Put an arc $a\to b$ for $a<b$ whenever the preceding auxiliary connectivity holds in $F[a,b]$. Suppose this acyclic digraph has no chain of $\ell$ vertices. Its longest-path ranks partition the clusters into at most $\ell-1$ classes.

For one rank class $L$, shortcut each row to consecutive clusters belonging to $L$, obtaining a multigraph $F_L$. Its edges expand to edge-disjoint paths in $F$, within their endpoint index intervals. No nontrivial subgraph of $F_L$ can be $2W$-edge-connected: its least and greatest indices would give an arc within one rank class. Recursively split at cuts below $2W$. This bounds $|E(F_L)|$ by $2W|L|$. If $T_L$ is the assigned-row mass in the class, then
\[
 T_L-N\le |E(F_L)|\le2W|L|\le T_L/2,
\]
because every cluster has at least $4W$ assignments. Thus $T_L\le2N$. Summing over the rank classes contradicts the total mass hypothesis. A chain of $\ell$ vertices exists and can be found by auxiliary max-flow tests followed by longest-path dynamic programming.

\paragraph{Disjoint connectors and endpoints.}
Process consecutive pairs of this chain in order. The appropriate endpoint sets have vertex connectivity at least $3u$. Before routing out of an internal cluster, delete the $u$ vertices used as incoming endpoints. At least $2u$ disjoint paths remain; choose $u$ and trim them at the endpoint clusters. Property (iii) excludes previous connector interiors, and endpoint deletion excludes the only possible remaining overlaps.

In the first cluster choose $u$ unused first endpoints for the external interface; in the last choose $u$ unused last endpoints. There are at least $4W$ distinct endpoints of each type, so this is possible. Every selected terminal vertex contributes an occurrence to the original endpoint multiset, and the selected terminals are distinct. They inherit full $\gamma$ edge-well-linkedness. All operations are polynomial-time computations.
\end{proof}

\begin{lemma}[Constructive strongification]\label{lem:strong}
A maximum-degree-$\Delta$ weak path-of-sets system of width $u$, whose terminal unions are fully $\gamma$ edge-well-linked, can be thinned in polynomial time to a strong system of the same length and width
\[
 \Omega\!\left(\frac{\gamma^3u}{\Delta^9r^2}\right),
\]
where $r=O(\log(2u))$ bounds the approximation factors in the constructive well-linkedness boosting theorem. Widths are assumed sufficiently large to absorb integer rounding. For fixed $\Delta$, the output of Lemma~\ref{lem:ordered} therefore has strong width $\Omega(\gamma^4W/r^2)$.
\end{lemma}
\begin{proof}
The algorithmic clause of Chekuri--Chuzhoy~\cite[Theorem~2.14]{CC16} extracts a node-well-linked subset of size at least $\gamma k/(32\Delta^4r)$ from a fully $\gamma$ edge-well-linked set of $k$ terminals. For each connector matching, recorded after trimming the physical paths, boost its left endpoints, and then boost their matched right endpoints. At least $\Omega(\gamma^2u/(\Delta^8r^2))$ pairs remain, and both endpoint sets are node-well-linked. Subsequent subsets preserve that property.

Temporarily retain a common number $k$ of pairs in every matching, and boost the external interfaces to the same size. Inside every cluster the two interfaces are individually node-well-linked and their union inherits full $\gamma$ edge-well-linkedness. The linked-subsets theorem~\cite[Theorem~2.9]{CC16} says that any equal-sized subsets of size at most $\gamma k/(2\Delta)$ are linked. Thin every matching to this common size. The connector paths remain disjoint, and the result is strong. The larger existential boosting bound quoted in~\cite{CT21} is not needed for this algorithmic argument.
\end{proof}

\section{Algorithms and running times}
\label{sec:algorithms}

We give explicit pseudocode for the two central routines and their common row consumer. The input graphs, row paths, designated intervals, and reference terminal sets are stored explicitly. Every contraction stores its branch set, so the output minor can be lifted. Integer ceilings in the logarithmic denominators make the flow capacities rational. Running times below count standard graph, linear-programming, and maximum-flow operations.

\subsection{The trimming and localization routines}

Algorithm~\ref{alg:trim} is the construction in Lemma~\ref{lem:flow-trimming}. The set reached from the source is taken in the residual network of a maximum flow, not in the original network. This distinction is what certifies all new boundary supplies after deletion.

\begin{algorithm}[tb]
\caption{\textsc{FlowTrim}$(C_0,\Gamma_0,\alpha_0,U)$}
\label{alg:trim}
\begin{algorithmic}[1]
\Require Fixed reference $C_0$ with fully $\alpha_0$-well-linked $\Gamma_0$; current $U\subseteq V(C_0)$.
\Ensure $U'\subseteq U$ as in Lemma~\ref{lem:flow-trimming}.
\State Create a directed network on $U\cup\{s,t\}$.
\For{$v\in U$}
  \State Add $s\to v$ with capacity $|E_{C_0}(\{v\},V(C_0)\setminus U)|$.
  \If{$v\in\Gamma_0$} \State Add $v\to t$ with capacity $\alpha_0/2$. \EndIf
\EndFor
\State Replace each edge of $C_0[U]$ by two opposite arcs of capacity one.
\State Compute a maximum $s$--$t$ flow and its residual source-reachable set $\{s\}\cup Z$.
\State $U'\gets U\setminus Z$.
\If{$\Gamma_0\cap U'=\varnothing$} \State \Return $\varnothing$. \EndIf
\State \Return the component of $C_0[U']$ containing all terminals in $\Gamma_0\cap U'$.
\end{algorithmic}
\end{algorithm}

For localization, each original cluster keeps a record $(C_0,\Gamma_0,U)$. Records never split into two descendants. A recursive cell may contain several records, and these are routed independently to its frontier. Vertex-disjoint linkage tests use vertex capacities, including capacity one at terminal and frontier vertices; a common vertex is allowed as a trivial path.

\begin{algorithm}[tb]
\caption{\textsc{Localize}$(J,\calR,\{(C_0,\Gamma_0)\},W)$}
\label{alg:localize}
\begin{algorithmic}[1]
\Require The input of Theorem~\ref{thm:flow-localization}.
\State Set $T_0,d,\delta$ as in that theorem; initialize every record with $U=V(C_0)$.
\State Put the full row cell, all records, and depth $0$ in a queue.
\While{the queue is nonempty}
  \State Remove a cell $I$ with its records and depth $j$.
  \State Sweep $I$ to a frontier $S$ balancing all its current terminals.
  \For{each record $(C_0,\Gamma_0,U)$ in $I$}
    \State $M\gets|\Gamma_0\cap U|$.
    \State Find a maximum vertex-disjoint linkage from $\Gamma_0\cap U$ to $S\cap U$ in $C_0[U]$.
    \If{its size is at least $\delta M$}
      \State Store $\lfloor\delta M\rfloor$ paths and their ports as a captured record at depth $j$.
    \Else
      \State Obtain a minimum vertex separator $K$ from the same flow computation.
      \State Let $D$ be the component of $C_0[U]-K$ containing more than $M/2$ terminals.
      \State $U\gets\Call{FlowTrim}{C_0,\Gamma_0,\alpha_0,V(D)}$.
      \State Place the updated record in the child cell on the side of $S$ containing $D$.
    \EndIf
  \EndFor
  \State Enqueue the nonempty child cells at depth $j+1$.
\EndWhile
\State Choose the depth of largest captured terminal mass.
\State \Return its ordered cells, captured clusters, and exported port sets.
\end{algorithmic}
\end{algorithm}

The component $D$ in Algorithm~\ref{alg:localize} exists by current full well-linkedness and the degree bound. Its identity is determined by counting current terminals in components. Lemma~\ref{lem:flow-trimming} restores the same reference guarantee after every call. The proof of Theorem~\ref{thm:flow-localization} shows that the empty return of \textsc{FlowTrim} cannot occur in this use, and that the queue has depth less than $d$.

\subsection{The common row consumer}

Both local dichotomies reduce to the same input: a degree-four graph
covered by perfect unique rows, together with disjoint paths that
meet many rows. Algorithm~\ref{alg:consumer} gives the shared consumer;
Section~\ref{sec:unbalanced-local} supplies its input.

The two decomposition stages use the following rule. In each current
induced piece, call the terminal-cut approximation and split only when
its returned ratio is below the specified threshold. Discard the
designations and endpoint occurrences of crossing intervals, but retain
the full induced host in each child. Source pruning deletes vertices
of the marked-incidence graph, not physical paths from the host.

\begin{algorithm}[tb]
\caption{\textsc{ConsumeRows}$(J,\mathcal Q,\mathcal S,w,\rho,\ell)$}
\label{alg:consumer}
\begin{algorithmic}[1]
\Require The degree-four core, unique rows $\mathcal Q$, and auxiliary paths $\mathcal S$ of Proposition~\ref{prop:consumer}; parameters from~\eqref{eq:explicit-budgets}.
\State Form the source clusters and their reference terminal sets by Lemma~\ref{lem:source}.
\State Let $T_0$ be the total terminal mass and set $d=\lceil\log_2(T_0+1)\rceil+2$; set $\delta,\beta,\gamma,W$ by~\eqref{eq:local-parameters} using the source guarantee $\alpha_0$.
\State Run \textsc{Localize}; expand its exported ports to whole row intervals by Lemma~\ref{lem:expansion}.
\State Build the row-labelled multigraph $F$ of Lemma~\ref{lem:ordered} and test every index interval by edge max flow.
\State Find a directed chain of $\ell$ clusters by longest-path dynamic programming; route the physical connectors in order, avoiding already used endpoints.
\State Apply constructive strongification and thin each interface to width $w$.
\State \Return the strong system in $J$ and its stored paths and interfaces.
\end{algorithmic}
\end{algorithm}

\paragraph{Running time and representation.}
Let the core have $n$ vertices and $e$ edges.
Each decomposition has $O(n)$ nonempty pieces, localization uses
$O(n\log n)$ maximum-flow calls, and ordered routing uses $O(n^2)$
connectivity tests and $O(n)$ physical linkage computations.
The flow networks have $O(n)$ vertices and $O(n+e)$ arcs.
Direct frontier maintenance takes $O(n(n+e)\log n)$ additional time.
The terminal-cut approximation and well-linkedness boosting are
polynomial-time routines, so the consumer has a uniform polynomial
running time.

The degree-four core stays fixed throughout the consumer.
Each trimming record keeps its original reference cluster and terminal
set; records never fork, and only children containing a record enter
the localization queue. Physical vertices and edges retain their
identities, including parallel edges. Every interval graph $J[a,b]$
is the exact physical edge union from Lemma~\ref{lem:ordered}, with
repeated edges included once. Connector interfaces are recorded after
trimming. Linkage computations use unit vertex capacities, including
terminals; the two occurrences of a singleton interval endpoint are
used only in edge-cut accounting. Contraction branch sets and
representative original edges give the final minor map.

\FloatBarrier
\section{One construction for both local dichotomies}
\label{sec:unbalanced-local}

The balanced and unequal-interface theorems use the same algorithm.
The only difference is how they reach the backbone paths that survive
the peeling step. We isolate that difference in the following elementary
flow observation.

\begin{lemma}[Retargeting a linkage]\label{lem:retarget-linkage}
Let $A,X$ be disjoint vertex sets, let $|A|=K\ge8M$, and let
$A'\subseteq A$ have size at least $K/2$. There is an $M$-linkage
from $X$ to $A'$ under either of the following hypotheses:
\begin{enumerate}[label=(\roman*),nosep]
\item there is a full $K$-linkage from $X$ to $A$;
\item $A$ is node-well-linked and there is a $2M$-linkage from $X$ to $A$.
\end{enumerate}
Such a linkage can be found by one vertex-capacitated flow computation.
\end{lemma}
\begin{proof}
In case (i), at least $|A'|\ge K/2\ge M$ paths of the given linkage
end in $A'$; retain any $M$ of them.
In case (ii), let $A_0$ be the $2M$ endpoints of the given linkage.
Since $|A'\setminus A_0|\ge K/2-2M\ge2M$, choose
$A_1\subseteq A'\setminus A_0$ of size $2M$.
Node-well-linkedness gives a $2M$-linkage from $A_0$ to $A_1$.
Concatenating the two linkages gives $2M$ walks from $X$ to $A'$
with vertex congestion at most two. Delete cycles and scale by one half.
This is a feasible value-$M$ flow with unit vertex capacities, so
integrality supplies the required linkage.
\end{proof}

\begin{theorem}[An unbalanced local dichotomy]
\label{thm:unbalanced-local}
There is an absolute constant $C$ with the following property.
Let $w,\rho,\ell\ge2$ be integers and put
$L=\lceil\log_2(w\rho\ell+2)\rceil$.
Suppose that $A,B,X$ are disjoint vertex sets in a graph $H$,
that $|A|=|B|=K$ and $|X|=h$, and that every vertex of $X$
has degree one. Assume that $A$ is node-well-linked, there is
a $K$-linkage from $A$ to $B$, and there is an $h$-linkage
from $X$ to $A$. If
\[
 K\ge Cw\rho\ell L^{50},
 \qquad h\ge Cw\rho L^{40},
\]
then a polynomial-time algorithm finds either an
$(A,B,X)$-crossbar of width $\rho$, or a degree-four minor
of $H$ containing a strong path-of-sets system of width $w$
and length $\ell$.
\end{theorem}

\begin{proof}[Proof of Theorems~\ref{thm:local} and~\ref{thm:unbalanced-local}]
Write $K=|A|=|B|$ and $h=|X|$; thus $h=K$ in the balanced case.
We first fix the backbone linkage and peel its paths.
Only afterwards do we choose the smaller linkage, which becomes
the unique row family in the consumer.

Apply Proposition~\ref{prop:consumer} with parameter $\rho_c=8\rho$,
and let $H_0,m$ be the resulting quantities in
\eqref{eq:explicit-budgets}. Put
\[
 M=512H_0\rho=64H_0\rho_c,
 \qquad D=\left\lfloor\frac K{2\rho}\right\rfloor.
\]
The stated bounds, after an absolute increase of their constants, give
\begin{equation}\label{eq:unbalanced-budgets}
 h\ge2M,\qquad K\ge8M,\qquad K\ge2m.
\end{equation}
For Theorem~\ref{thm:local}, take $c_0=50$.
Indeed, replacing $\rho$ by $8\rho$ increases the logarithm by at most
three; in the balanced case $h=K$, and its single size bound implies
all three inequalities. No rounding of $\rho$ is needed.

\paragraph{Peel the backbone.}
Fix a full $A$--$B$ linkage $\mathcal P$.
Its paths avoid the leaves $X$.
Initially every path is available. In iteration $i\in[D]$, form a
fresh quotient of the original graph $H$ by contracting each
available path to a separate vertex. Compute a maximum
vertex-disjoint linkage from these contracted vertices to $X$.

If its value is at least $\rho$, orient $\rho$ flow paths from $X$
and truncate each at the first contracted vertex it meets.
Their contracted endpoints are distinct and their interiors avoid
all available backbone paths. Lift them using the original incident
edges. Together with their $\rho$ backbone paths, they are the
required crossbar.

Otherwise take a vertex separator $J_i$ of size less than $\rho$.
Let $\mathcal R_i$ be the available paths whose contracted vertices
belong to $J_i$, and let $J_i''$ be its ordinary vertices.
Mark the paths of $\mathcal R_i$ unavailable. In the original graph,
\[
 V_i=J_i''\cup\bigcup_{P\in\mathcal R_i}V(P)
\]
separates $X$ from every path still available.
This is the separator step of the pseudo-grid construction
in~\cite[Theorem~4.1]{CT21}, applied before choosing an auxiliary linkage.

If all iterations finish, the families $\mathcal R_i$ are disjoint and
each contains fewer than $\rho$ paths. Therefore
\[
 \mathcal R=\bigcup_i\mathcal R_i,
 \qquad |\mathcal R|\le\rho D\le K/2.
\]
Let $A_{\rm rem}$ consist of the $A$ endpoints of the available paths;
then $|A_{\rm rem}|\ge K/2$.

\paragraph{Choose the smaller linkage and make it unique.}
Lemma~\ref{lem:retarget-linkage} supplies an $M$-linkage from $X$
to $A_{\rm rem}$: use case (i) in the balanced theorem and case (ii)
in the unequal-interface theorem. This is the only difference between
the two applications. The lemma certifies feasibility for the following
minimum-cost flow; no separate retargeting paths are computed.

Apply Lemma~\ref{lem:mincost-linkage} with fixed path graph
$F=\bigcup_{P\in\mathcal P}P$, source set $X$, target set
$A_{\rm rem}$, and flow value $M$. It constructs a minimum-cost
linkage $\mathcal Q$ whose edges outside $F$ are essential in
$F\cup\mathcal Q$, also after fixing its selected endpoint sets.

Every $Q\in\mathcal Q$ meets $V_i$ for each $i$.
Fewer than $\rho$ of these disjoint paths can meet $J_i''$.
All the others meet a path of $\mathcal R_i$.
Consequently the number $I$ of distinct intersecting pairs
in $\mathcal R\times\mathcal Q$ satisfies
\begin{equation}\label{eq:dual-incidence}
 I\ge(M-\rho)D.
\end{equation}
Pairs charged at different iterations are distinct because the
$\mathcal R_i$ are disjoint. We need only this total count, so there
is no need to discard paths exceptional in some iteration.

Apply Lemma~\ref{lem:essential-core} with fixed path graph $F$
and row linkage $\mathcal Q$. Its selected source endpoints lie in $X$,
which is disjoint from $F$. It gives a degree-four minor covered
by the $M$ images of $\mathcal Q$, which form a perfect unique
linkage. Every incidence with a backbone path is preserved, and the
nonempty backbone images remain mutually vertex-disjoint paths.
Continue to use the same notation for these images; empty images
had no incidence and can be discarded.

\paragraph{Cut the backbone images into auxiliary segments.}
Only after the core contractions, traverse each $P\in\mathcal R$.
End a segment as soon as it has met $64H_0$ distinct rows of
$\mathcal Q$, and start the next segment at the next vertex.
If $P$ meets $d(P)$ distinct rows, the number $t(P)$ of complete
segments satisfies
\[
 d(P)\le64H_0t(P)+(64H_0-1),
 \qquad t(P)\ge\frac{d(P)}{64H_0}-1.
\]
This uses only the number of distinct rows per segment, and allows
a row to be visited repeatedly. Since $|\mathcal R|\le\rho D$,
the number of complete segments is at least
\[
 \frac I{64H_0}-|\mathcal R|
 \ge\left(\frac{M-\rho}{64H_0}-\rho\right)D
 \ge6\rho D\ge m.
\]
The last inequality follows from $D\ge K/(4\rho)$ and $K\ge2m$;
the former holds because \eqref{eq:unbalanced-budgets} implies
$K\ge8M\ge4\rho$.
All these segments are vertex-disjoint. Retain any $m$ of them.
Together with the $M=64H_0\rho_c$ unique rows, they satisfy
Proposition~\ref{prop:consumer}, which returns the required strong system.

There are at most $K$ peeling flows and one minimum-cost flow.
The retargeting lemma proves feasibility for the particular
$A_{\rm rem}$ encountered; the algorithm does not test a condition
over all subsets of $A$.
The core contractions, segmentation, and consumer are polynomial-time
operations. Recording their contraction branch sets gives the minor
map into the original $H$ and proves both theorems.
\end{proof}

\begin{algorithm}[tb]
\caption{\textsc{LocalDichotomy}$(H,A,B,X,w,\rho,\ell)$}
\label{alg:local}\label{alg:unbalanced-local}
\begin{algorithmic}[1]
\Require The hypotheses of Theorem~\ref{thm:local} or Theorem~\ref{thm:unbalanced-local}.
\State Set $\rho_c=8\rho$; take $H_0,m$ from Proposition~\ref{prop:consumer}; set $M=512H_0\rho$ and $D=\lfloor |A|/(2\rho)\rfloor$.
\State Find a full $A$--$B$ linkage $\mathcal P$; mark all its paths available and set $\mathcal R=\varnothing$.
\For{$i=1,\ldots,D$}
  \State Contract every available path to one vertex in a fresh copy of the original $H$.
  \State Compute a maximum vertex flow from the contracted vertices to $X$.
  \If{the flow value is at least $\rho$}
    \State Orient $\rho$ flow paths from $X$, truncate at their first contracted vertices, lift them, and \Return the resulting crossbar.
  \EndIf
  \State Take a minimum vertex separator $J_i$; let $\mathcal R_i$ be the available paths whose contracted vertices belong to $J_i$.
  \State Set $\mathcal R\gets\mathcal R\cup\mathcal R_i$ and mark the paths in $\mathcal R_i$ unavailable.
\EndFor
\State Let $A_{\rm rem}$ be the $A$ endpoints of the available paths.
\State Compute a minimum-cost vertex flow of value $M$ from $X$ to $A_{\rm rem}$, with edge costs zero on $\mathcal P$ and one elsewhere; delete cycles to obtain $\mathcal Q$.
\State Apply Lemma~\ref{lem:essential-core} to $\mathcal P\cup\mathcal Q$, with fixed path graph $\mathcal P$ and row linkage $\mathcal Q$; record its minor map.
\State Split the surviving images of paths in $\mathcal R$ into vertex-disjoint segments, ending each at its $64H_0$th distinct $\mathcal Q$ row; retain $m$ segments.
\State Apply \textsc{ConsumeRows} with parameters $w,\rho_c,\ell$ to the rows and retained segments.
\State \Return its strong system and the composed minor map into $H$.
\end{algorithmic}
\end{algorithm}

\FloatBarrier
\section{Weighted separators and grid construction}
\label{sec:weighted-separators}

When a vertex represents a contracted cluster, its full boundary is the natural cost of deleting it. The terminals relevant to a particular application may form only part of that boundary. We therefore keep separator cost and terminal mass separate. We first prove the one-weight construction, including the algorithm that finds the grid. The argument follows the separator-or-minor method of Krivelevich and Nenadov~\cite[Theorem~8.1]{Krivelevich19}.

For disjoint vertex sets $U,C$ or for $U\subseteq C$, write $N_C(U)$ for the vertices of $C\setminus U$ having a neighbor in $U$. Parallel edges can be suppressed throughout the algorithms in this section; this does not change vertex separators or grid-minor models. They are retained when applying the edge-well-linkedness corollary.

\begin{lemma}[The weighted grid primitive]
\label{lem:weighted-grid-primitive}
Let $F$ have positive integral vertex costs $b(v)\le R$, total cost $m$, and let $L_m=\lceil\log_2(2m)\rceil$. Given a rational $0<a\le1/16$ and an integer $g\ge2$ satisfying
\begin{equation}\label{eq:weighted-kn-budget}
 g^2\le\frac{a^2m}{64RL_m},
\end{equation}
a polynomial-time algorithm returns either a $g\times g$ grid minor or a vertex set of cost at most $am$ whose deletion leaves components of cost at most $2m/3$ each.
\end{lemma}

\begin{proof}
We first describe the graph-search subroutine used by the construction.

\paragraph{A sparse ball or short paths.}
In an induced subgraph $F[C]$, a deterministic subroutine either finds a nonempty $U\subseteq C$ with
\[
 b(U)\le b(C)/2,\qquad b(N_C(U))\le a b(U),
\]
or certifies that $F[C]$ is connected and has diameter at most
\[
 D=2\left\lceil\frac{2L_m}{a}\right\rceil.
\]
Run breadth-first search from each vertex of $C$ and inspect its nonempty balls of cost at most $b(C)/2$. Stop each search before testing the first ball of larger cost. Return any eligible ball satisfying the neighborhood inequality. If a connected component has cost at most $b(C)/2$, its final ball qualifies. Thus failure to find one implies that $F[C]$ is connected.

Put $h=\lceil2L_m/a\rceil$. Until a ball has cost greater than $b(C)/2$, its cost grows by a factor greater than $1+a$ at each step. Its initial cost is at least one. Since $\ln(1+a)\ge a/2$ and $L_m\ge\ln(2m)$, a ball of radius $h$ must have cost greater than $b(C)/2$: otherwise its cost would exceed $(1+a)^h\ge\exp(L_m)\ge2m$. Two such balls cannot be disjoint, so every pair of vertices has distance at most $2h=D$. The subroutine scans only actual breadth-first-search layers, at most $|C|$ for each starting vertex; it does not take a number of steps proportional to $1/a$.

\paragraph{Building and discarding branch sets.}
Index the target grid by $[g]\times[g]$. Maintain a partition $V(F)=A\mathbin{\dot\cup}B\mathbin{\dot\cup}C$, where $A$ is discarded, $C$ is unused, and $B$ is the disjoint union of connected branch sets $B_x$ indexed by $I\subseteq[g]\times[g]$. These branch sets model the grid induced by $I$. Maintain
\begin{equation}\label{eq:weighted-buffer-invariants}
 b(N_C(A))\le a b(A),\qquad
 b(B_x)\le R(3D+1),\qquad b(B)\le am/4.
\end{equation}
Initially $A=B=\varnothing$, $C=V(F)$, and $I=\varnothing$.

While $b(A)<m/3$ and $I$ is incomplete, select a missing grid vertex $x$. It has at most four grid neighbors in $I$. If some such neighbor $y$ has $N_C(B_y)=\varnothing$, move $B_y$ from $B$ to $A$ and remove $y$ from $I$. This rollback preserves all required grid adjacencies among the remaining branch sets and preserves the first invariant, since $B_y$ has no neighbor in $C$.

Otherwise run the sparse-ball subroutine on $F[C]$. If it finds $U$, move $U$ from $C$ to $A$. The neighborhood invariant follows from
\[
 N_{C\setminus U}(A\cup U)\subseteq N_C(A)\cup N_C(U).
\]
If it certifies diameter at most $D$, choose $v_y\in N_C(B_y)$ for each already represented grid neighbor $y$ of $x$. Join these at most four vertices inside $C$ by at most three shortest paths from one chosen pivot. Their union $Y$ is connected and has at most $3D+1$ vertices. If there is no represented neighbor, take any one vertex of $C$. Set $B_x=Y$, move $Y$ from $C$ to $B$, and add $x$ to $I$. The required adjacencies are present, and deleting vertices from $C$ preserves the neighborhood invariant.

Since $a\le1/16$ and $L_m\ge1$,
\[
 D\le5L_m/a,\qquad 3D+1\le16L_m/a.
\]
There are at most $g^2$ branch sets, so~\eqref{eq:weighted-kn-budget} gives
\[
 b(B)\le g^2R(3D+1)\le\frac{16g^2RL_m}{a}\le\frac{am}{4}.
\]
Before termination, $b(A)<m/3$ and $b(B)\le am/4$, so $C$ is nonempty and every operation is defined.

If $I$ becomes complete, return its grid model. Otherwise consider the first time $b(A)\ge m/3$. A sparse-ball deletion adds at most $b(C)/2$, leaving $b(A)<2m/3$. A rollback adds at most the preceding cost of $B$, which is at most $am/4\le m/64$, and also leaves $b(A)<2m/3$. Put $Z=B\cup N_C(A)$. Then
\[
 b(Z)\le am/4+a b(A)\le11am/12\le am.
\]
There is no edge from $A$ to $C\setminus N_C(A)$. Their costs are at most $2m/3$ each, because $m/3\le b(A)\le2m/3$. Every component of $F-Z$ therefore has cost at most $2m/3$.

Each sparse-ball deletion or insertion removes at least one vertex from $C$, and no vertex returns to $C$. Each rollback removes a branch set created by an earlier insertion. There are at most $|V(F)|$ deletions and insertions combined and at most $|V(F)|$ rollbacks. Every iteration uses polynomially many graph searches and exact rational comparisons. The running time is polynomial, independently of the value of $1/a$.
\end{proof}

\begin{algorithm}[tb]
\caption{\textsc{WeightedGridOrSeparator}$(F,b,R,g,a)$}
\label{alg:weighted-grid-primitive}
\begin{algorithmic}[1]
\Require The hypotheses of Lemma~\ref{lem:weighted-grid-primitive}.
\State Set $A=B=\varnothing$, $C=V(F)$, $I=\varnothing$; let $Q$ be the $g\times g$ grid.
\While{$b(A)<b(V(F))/3$ and $I\ne V(Q)$}
  \State Choose $x\in V(Q)\setminus I$; put $X=N_Q(x)\cap I$.
  \If{some $y\in X$ has $N_C(B_y)=\varnothing$}
    \State Move $B_y$ from $B$ to $A$; remove $y$ from $I$; \textbf{continue}.
  \EndIf
  \State Run the sparse-ball subroutine in $F[C]$.
  \If{it returns a sparse set $U$}
    \State Move $U$ from $C$ to $A$; \textbf{continue}.
  \EndIf
  \State Choose $v_y\in N_C(B_y)$ for each $y\in X$.
  \State Join these vertices by shortest paths from one pivot in $F[C]$; if $X=\varnothing$, take one vertex of $C$. Call the resulting connected set $Y$.
  \State Set $B_x=Y$; move $Y$ from $C$ to $B$; add $x$ to $I$.
\EndWhile
\If{$I=V(Q)$}
  \State \Return the grid model $(B_x)_{x\in V(Q)}$.
\Else
  \State \Return the separator $B\cup N_C(A)$.
\EndIf
\end{algorithmic}
\end{algorithm}

\FloatBarrier
\begin{lemma}[Two-weight separator or grid]\label{lem:two-weight-kn}
There is an absolute constant $C$ and a polynomial-time algorithm with the following guarantee. Let $J$ be a graph, possibly with parallel edges, let $b:V(J)\to\mathbb Z_{>0}$ be costs, and let $\tau:V(J)\to\mathbb Z_{\ge0}$ be terminal weights. For a specified set $D$ to delete initially, put
\[
 K=\tau(V(J)),\quad M=b(V(J)\setminus D),\quad
 d=b(D),\quad R=\max_{v\notin D}b(v).
\]
Assume $K,M>0$, and let $g\ge2$ be an integer. The algorithm returns a $g\times g$ grid minor in $J-D$, or a set $Z\supseteq D$ such that every component of $J-Z$ has terminal weight at most $2K/3$, and
\begin{equation}\label{eq:two-weight-separator-cost}
 b(Z)\le d+Cg\sqrt{RM\log_2(2M)}.
\end{equation}
The bound $R$ is required only outside $D$. One may take $C=2^{11}$.
\end{lemma}
\begin{proof}
Put $L=\lceil\log_2(2M)\rceil$ and start with $S=\varnothing$. If every component of $J-D-S$ has terminal weight at most $2K/3$, return $Z=D\cup S$. Otherwise there is exactly one component $C_i$ of terminal weight greater than $2K/3$. Write $m_i=b(C_i)$ and, for the analysis, $x_i=8g\sqrt{RL/m_i}$.

If $x_i>1/16$, add all of $C_i$ to $S$ and terminate. This costs $m_i<128g\sqrt{Rm_iL}$. Otherwise choose the smallest dyadic $a_i=2^{-j}\ge x_i$. Then $x_i\le a_i<2x_i$ and $a_i\le1/16$. These choices require only integer comparisons: the first case is $m_i<2^{14}g^2RL$; otherwise $j\ge4$ is the largest integer with $2^{2j+6}g^2RL\le m_i$. No square root is computed.

Because $\lceil\log_2(2m_i)\rceil\le L$, the budget of Lemma~\ref{lem:weighted-grid-primitive} holds. Apply it to $J[C_i]$. Return any grid found. Otherwise its separator $S_i$ costs at most $a_i m_i<16g\sqrt{Rm_iL}$ and leaves components of cost at most $2m_i/3$. Add $S_i$ to $S$ and repeat. The next terminal-heavy component, if any, lies inside $C_i$, so $m_{i+1}\le2m_i/3$.

All separators and the possible final deletion cost at most
\[
 128g\sqrt{RL}\sum_{i\ge0}\sqrt{M(2/3)^i}
 =\frac{128}{1-\sqrt{2/3}}g\sqrt{RML}
 \le2^{11}g\sqrt{RM\log_2(2M)}.
\]
Here $L\le2\log_2(2M)$ and $(1-\sqrt{2/3})^{-1}<6$. There are $O(\log(2M))$ calls and the dyadic exponents are $O(\log(2M))$. All weights in our quotient applications are boundary counts, with $M=O(n^2)$, so the total running time is polynomial in the graph order and $g$.
\end{proof}

\begin{corollary}[A well-linked quotient with prescribed deletions]
\label{cor:robust-quotient-budget}
Let $F$ have positive integral vertex costs $b$ and nonnegative integral
terminal weights $\tau$, with $\deg_F(v),\tau(v)\le b(v)$.
Suppose $K=\tau(V(F))>0$ and $\tau$ is fully $\alpha$ edge-well-linked
in $F$, where $0<\alpha\le1$. For a prescribed set $D$, put
$d=b(D)$ and $M=b(V(F)\setminus D)$. Let $R\ge1$ bound all costs
outside $D$, and let $\mathcal L\ge\log_2(2\max\{1,M\})$.
For every integer $g\ge2$, a polynomial-time algorithm returns either
a $g\times g$ grid minor in $F-D$, or a set $Z\supseteq D$ such that
every component of $F-Z$ has terminal weight at most $2K/3$ and
\begin{equation}\label{eq:robust-quotient-budget}
 \frac{\alpha K}{6}\le b(Z)\le d+Cg\sqrt{RM\mathcal L},
\end{equation}
where $C$ is the constant of Lemma~\ref{lem:two-weight-kn}.
In particular, a grid is returned whenever the upper bound in
\eqref{eq:robust-quotient-budget} is smaller than $\alpha K/6$.
The costs of vertices in $D$ need not be bounded by $R$.
\end{corollary}
\begin{proof}
If $M=0$, return $Z=D=V(F)$; here $K\le b(D)$.
Otherwise apply Lemma~\ref{lem:two-weight-kn}. Only the lower
separator bound needs proof. It is immediate when $b(Z)\ge K/6$.
In the remaining case, $\tau(Z)<K/6$, and the components of $F-Z$
have total terminal weight greater than $5K/6$, with each at most
$2K/3$. Some union $U$ of components therefore has weight in
$[K/6,2K/3]$: use one component in this interval, or add smaller
components until first reaching $K/6$. Every edge leaving $U$ meets
$Z$, so
\[
 \alpha K/6\le |E_F(U,V(F)\setminus U)|
 \le\sum_{v\in Z}\deg_F(v)\le b(Z).
\]
\end{proof}

\paragraph{Quotient models.}
Every contracted piece is connected, and quotient cuts preserve
crossing edges and designated terminal mass. A piece's full boundary
is a valid deletion cost even when it includes edges leaving the
current host: it dominates both quotient degree and terminal
multiplicity. The prescribed set $D$ contains the virtual hub or
outside pieces that the grid must avoid. Deleting $D$ before the
search ensures that the returned model lifts using only the
remaining pieces.

\section{An accessible small-boundary router or a grid}\label{sec:hierarchy}

For a vertex set $S$ in a graph $J$, write $b_J(S)=|\delta_J(S)|$.
Write $\Gamma_J(S)$ for the set of vertices in $S$ incident to an edge of $\delta_J(S)$. Full-boundary terminals are incidences: a vertex incident to several boundary edges carries the corresponding number of terminal occurrences. Thus $\Gamma_J(S)$ denotes their support, not their multiplicities.

\begin{lemma}[A constructive routing hierarchy]\label{lem:bkr-hierarchy}
Let $J$ be a connected unweighted graph on $n$ vertices. In polynomial time one can construct a laminar hierarchy whose root is $V(J)$ and whose leaves are the singleton vertex sets, with the following properties. Every cluster induces a connected subgraph; every child has at most two thirds as many vertices as its parent; and the hierarchy has height $d=O(\log n)$. There are parameters
\[
\beta=\Omega(\log^{-2}(2n)),\qquad \theta=\Omega(\log^{-3}(2n))
\]
such that the full boundary of every cluster $S$ is $\beta$ edge-well-linked in $J[S]$. Moreover, let $S_1,\ldots,S_t$ be the children of a non-leaf cluster $S$, and let $w_S(v)$ count the edges incident to $v$ that leave its child, including edges leaving $S$ itself. The demand
\[
\frac{w_S(u)w_S(v)}{w_S(S)}
\]
for every ordered pair $(u,v)$ can be routed simultaneously in $J[S]$ with congestion at most $1/\theta$. The routings can be computed in polynomial time.
\end{lemma}

\begin{proof}
We give a direct implementation of the published construction of Bie\'nkowski, Korzeniowski, and R\"acke~\cite{BKR03}. For $U\subseteq S$, write $b_S(U)=|E_J(U,V(J)\setminus S)|$ and $b(S)=b_S(S)$. Let $L=\lceil\log_2(2n)\rceil$. By the deterministic cut algorithm of Linial, London, and Rabinovich~\cite[Theorem~4.1]{LLR95}, we may fix an absolute constant $c_{\rm FC}$ and take $\sigma=\lceil c_{\rm FC}L\rceil$ so that its returned cut has sparsity at most $\sigma$ times the optimal concurrent-flow throughput for every demand instance below. Put
\[
 \lambda=64\sigma L,\qquad \varepsilon=4\sigma/\lambda=1/(16L),\qquad
 \beta=1/\lambda,\qquad \theta=1/(24\sigma\lambda).
\]
We maintain the precondition
\begin{equation}\label{eq:bkr-explicit-precondition}
 \lambda |E_J(U,S\setminus U)|\ge b_S(U)
 \qquad(U\subseteq S,\ |U|\le3|S|/4).
\end{equation}
\paragraph{Refinement.}
For any nonempty set $R$, start with the partition $\{R\}$. For a current part $P$, apply the cut algorithm in $J[P]$ to the demands $d_{uv}=b_P(\{u\})/|P|$ for ordered pairs $u\ne v$. For its returned cut $P=X\mathbin{\dot\cup}Y$, put
\[
 e=|E_J(X,Y)|,\qquad
 D=\frac{|Y|b_P(X)+|X|b_P(Y)}{|P|}.
\]
Split $P$ when $e/D\le\varepsilon$ and queue the two new parts; otherwise mark $P$ final. Singletons and zero-boundary parts are final without a test; a zero-capacity cut with positive demand is handled directly. There are only $O(|R|)$ cut calls. Every final part satisfies~\eqref{eq:bkr-explicit-precondition}: a violating $U$ has crossing demand at least $b_P(U)/4$, so the cut algorithm would return sparsity less than $4\sigma/\lambda$.

The final boundary sum is at most $2b(R)$. To see this, use $\Psi=\sum_P b(P)\log_2|P|$ over the current partition. Writing $p=|P|$, $x=|X|$, and $y=|Y|$, a split gives
\[
\begin{aligned}
 \Psi_{\rm before}-\Psi_{\rm after}
 &=b_P(X)\log_2(p/x)+b_P(Y)\log_2(p/y)
   -e(\log_2x+\log_2y)\\
 &\ge D-2eL\ge7D/8.
\end{aligned}
\]
Its boundary increase is $2e\le(16\varepsilon/7)(\Psi_{\rm before}-\Psi_{\rm after})$. Summing, and using $\Psi\ge0$, bounds the total increase by $b(R)/7$. Denote this procedure by $\mathsf{Refine}(R)$.

\paragraph{Partitioning one cluster.}
Let $S$ satisfy~\eqref{eq:bkr-explicit-precondition}, with $|S|\ge2$. Start with singleton children. For the current children $C$, put $w(v)=b_C(\{v\})$ when $v\in C$, extend $w$ additively, and put $W=w(S)$ and $d_{uv}=w(u)w(v)/W$. Test whether throughput $\theta$ is feasible in $J[S]$. For each commodity $(u,v)$ and each physical edge $e=\{a,b\}$, use nonnegative variables $f^{uv}_{e,a\to b}$ and $f^{uv}_{e,b\to a}$, with conservation and shared capacities
\[
 \operatorname{div}f^{uv}(x)
   =\theta d_{uv}(\mathbf1_{x=u}-\mathbf1_{x=v}),\qquad
 \sum_{u\ne v}(f^{uv}_{e,a\to b}+f^{uv}_{e,b\to a})\le1.
\]
Diagonal demands are fulfilled locally. If feasible, return the partition and scale the flow by $1/\theta$, obtaining the full demands with congestion $1/\theta$.

Otherwise the cut algorithm gives a cut $(A,B)$, oriented so that $|A|\le|S|/2$, whose size $c$ satisfies $c\le w(A)/(12\lambda)$: its crossing demand is $2w(A)w(B)/W\le2w(A)$. Let $U$ be the union of children $C$ with $|A\cap C|\ge3|C|/4$. Then $|U|\le2|S|/3$. For each child put $c_C=|E_J(A\cap C,C\setminus A)|$. Every mismatched piece $Q$ in $A\triangle U$ has at most three quarters of its child's vertices, so $b(Q)\le(\lambda+1)c_C$ by the precondition. Since $\sum_Cc_C\le c$, cut subadditivity gives
\[
 b(U)\le b(A)+\sum_Qb(Q)\le2(\lambda+1)c,\qquad
 w(U)\ge w(A)-\lambda c\ge11w(A)/12>0.
\]
Thus $U\ne\varnothing$ and $w(U)\ge3b(U)$, as $\lambda\ge2$.

Replace its children by $\mathsf{Refine}(U)$. This preserves the precondition and the two-thirds size bound. The integer potential
\[
 \Phi=\sum_{C\text{ a child}}b(C)-b(S)
\]
starts at $2|E(J[S])|$ and decreases by at least $3b(U)-2b(U)=b(U)\ge1$ at each replacement. The last inequality uses connectedness of $J$ and $0<|U|<n$. Consequently the procedure terminates with the required product-demand routing after polynomially many iterations.

\paragraph{The hierarchy.}
The root $V(J)$ satisfies the precondition because its boundary is empty. Recursively apply the partition procedure until reaching singletons. Every child has at most two thirds of its parent's vertices, giving height $O(\log n)$. The precondition implies that every cluster is connected: a smallest component of a disconnected cluster has zero internal boundary, so its external boundary would also vanish, contradicting connectedness of $J$. Applying the precondition to the cardinality-smaller side of a cut proves full boundary $\beta$ edge-well-linkedness. The weight $w(v)$ counts all edges leaving its child, including edges leaving $S$ itself, exactly as required.

There are polynomially many clusters and improvements. All physical edge capacities remain one, and all boundary counts and demand coefficients have polynomial encoding length. The displayed flow programs have $O(n^2|E(J)|)$ variables; all remaining tests use the deterministic cut algorithm. Thus the total running time is polynomial, with $\beta=\Omega(L^{-2})$ and $\theta=\Omega(L^{-3})$.
\end{proof}

\begin{lemma}[Expansion of a child quotient]\label{lem:bkr-child-quotient}
Contract the children $S_i$ of a hierarchy cluster $S$ in $J[S]$, and assign quotient vertex $i$ weight $b_i=b_J(S_i)$. Put $M=\sum_i b_i$ and $\eta=\theta/2$. Then
\[
\deg(i)\le b_i,\qquad M\ge b_J(S),
\]
and every quotient cut has at least
\[
\eta\min\left\{\sum_{i\in I}b_i,\sum_{i\notin I}b_i\right\}
\]
crossing edges. Parallel edges are counted with multiplicity.
\end{lemma}
\begin{proof}
Every edge leaving a child is either an edge of the quotient or an edge leaving $S$, proving the first two assertions. For a union of children, the product demand crossing the cut in one direction is $b(I)(M-b(I))/M$. Its routing with congestion $1/\theta$ gives a cut of size at least $\theta b(I)(M-b(I))/M\ge(\theta/2)\min\{b(I),M-b(I)\}$.
\end{proof}

\begin{lemma}[Access along the hierarchy]\label{lem:hierarchy-access}
Let $z$ be a vertex of $J$, and let $Y$ be a non-singleton hierarchy cluster with $z\notin Y$. Suppose every hierarchy edge on the path from $Y$ to the leaf $\{z\}$ has boundary capacity at least $r$. Any $r$ selected boundary incidences of $Y$ can send one unit each to $z$ with edge congestion at most $\lceil4(d+1)/\theta\rceil$.
\end{lemma}
\begin{proof}
Associate a cluster $S$ of boundary at least $r$ with the distribution placing $r b_S(v)/b_J(S)$ units at each vertex $v\in S$, where $b_S(v)$ is its boundary incidence count. Two distributions $x,y$ of total mass $r$, both coordinatewise bounded by $w_S$, can be routed to each other inside $S$ with congestion $2/\theta$: first send $x$ to the distribution $r w_S/w_S(S)$, then send that distribution to $y$, using the product flow. Each of the two demand matrices is coordinatewise bounded by the product demand.

Start by routing the selected boundary supplies to the boundary distribution of $Y$ inside $Y$. Both vectors are bounded by $w_Y$. Move upward toward the least common ancestor of $Y$ and $\{z\}$, changing the distribution of a child to that of its parent inside the parent only while that parent lies strictly below the least common ancestor. At the least common ancestor, change directly between the two child distributions; its own boundary distribution is never used. Then move downward to $\{z\}$. Each transition is dominated by the relevant parent's full child-boundary weight, because all used boundary capacities are at least $r$. At most $2d+2$ transitions are needed. Summing their congestion bounds proves the assertion.

Only the existence of these routes is used below. The attachment paths themselves will be obtained by one unit-capacity maximum flow in the original graph, after the surviving terminal set has been chosen.
\end{proof}

\begin{theorem}[Accessible router or grid]\label{thm:accessible-router-or-grid}
Let $G$ be a connected graph of maximum degree $\Delta$, and let $T$ be an $\alpha_0$ edge-well-linked set of $k$ vertices, where $0<\alpha_0\le1$. Put $B=\alpha_0 k/100$, and choose a parameter $R$ with $\Delta+1<R<B$. There is a polynomial-time algorithm that finds either
\begin{enumerate}
\item a grid minor of side
\[
\Omega\!\left(
\frac{\alpha_0\beta\theta}{d+1}
\sqrt{\frac{k}{R\log(2n)}}\right),
\]
where $n=|V(G)|+1$ and $d,\beta,\theta$ are the hierarchy parameters; or
\item a connected vertex set $Y\subseteq V(G)$ such that, in $J=G+\{zt:t\in T\}$,
\[
R\le b_J(Y)<B.
\]
The full boundary of $Y$ in $J$ is $\beta$ edge-well-linked in $G[Y]$, and all boundary incidences can route to $z$ with congestion $O((d+1)/\theta)$. Its actual boundary in $G$ has size at least $\alpha_0 b_J(Y)/(1+\alpha_0)$.
\end{enumerate}
All graphs in the grid outcome are minors of $G$; in particular, no branch set of the returned model contains the artificial vertex $z$.
\end{theorem}

\begin{proof}
Fix a target integer $g\ge2$ below a sufficiently small absolute constant times the claimed grid bound. Apply Lemma~\ref{lem:bkr-hierarchy} to $J$, and give the hierarchy edge from a non-root cluster $S$ to its parent capacity $b_J(S)$. Put
\[
 \eta=\theta/2,\qquad \mathcal L=\lceil\log_2(4n^2)\rceil,\qquad t=6Cg\sqrt{R\mathcal L}.
\]
The degree bound and $\Delta<R<B\le k\le n$ give $2|E(J)|\le2n^2$. Every sum of full boundary costs over disjoint clusters is at most $2|E(J)|$. Thus $\mathcal L$ bounds all logarithms required below, and $\mathcal L=O(\log(2n))$.

Let $\mathcal K$ be the component containing the leaf $\{z\}$ after retaining only hierarchy edges of capacity at least $B$. The edge incident to that leaf has capacity $k$, so it is retained. First look for a child $Y$ of a node in $\mathcal K$ such that $z\notin Y$ and $R\le b_J(Y)<B$. Every hierarchy edge from $Y$ to $\{z\}$ has capacity at least $b_J(Y)$. Lemma~\ref{lem:hierarchy-access} therefore routes all boundary incidences of $Y$ to $z$, and the hierarchy supplies full boundary well-linkedness in $G[Y]$. Here $Y$ is not a singleton, since $\Delta+1<R$. Moreover, $|T\cap Y|\le b_J(Y)<k/2$, and hence
\[
 b_G(Y)\ge\alpha_0|T\cap Y|,\qquad b_J(Y)=b_G(Y)+|T\cap Y|.
\]
These inequalities give the required actual-boundary bound.

Suppose no such child exists. If $\mathcal K$ contains a cluster not containing $z$, choose a deepest one, say $S$. Its boundary is at least $B$, while every child has boundary less than $R$: a child of boundary at least $B$ would contradict the choice of $S$, and one of boundary in $[R,B)$ would have been found. The child quotient has full-boundary expansion $\eta$, maximum vertex cost less than $R$, and total cost at least $B$. Corollary~\ref{cor:robust-quotient-budget}, with no prescribed deletion, constructs a grid of side $\Omega(\eta\sqrt{B/(R\mathcal L)})$. This is at least the target $g$ when its absolute constant is chosen sufficiently small. All branch sets lie in $G[S]$.

It remains that $\mathcal K$ is a chain
\[
 S_0\supset S_1\supset\cdots\supset S_s=\{z\},\qquad 1\le s\le d.
\]
For $i<s$, call the children of $S_i$ other than $S_{i+1}$ its side clusters. They all have boundary less than $R$; together they partition $S_0\setminus\{z\}$ into connected sets. Write
\[
 c_i=b_J(S_i),\qquad C_* =\max_i c_i,\qquad
 m_i=\sum_{\substack{U\text{ a side}\\\text{cluster of }S_i}}b_J(U),\qquad M=\sum_{i<s}m_i.
\]
Choose $j$ with $c_j=C_*$. We have $C_*\ge k$ and $d_0:=c_0<B$, with $d_0=0$ when $S_0$ is the root.

We apply the same quotient algorithm to the following graphs. For $Q_T$, work in $G$, contracting all side clusters and each connected component outside $S_0$ separately; use the original terminals $T$ and delete the outside-component vertices. For $Q_*$, work in $J[S_j]$, contract all side clusters contained in $S_j$, including those from deeper chain levels, use the full boundary of $S_j$ as terminals, and delete $z$. For each $Q_i$, contract the children of $S_i$ in $J[S_i]$, use all their full boundaries as terminal weights, and delete the child $S_{i+1}$.

Give every side-cluster vertex its full boundary cost in $J$. The exceptional vertex in $Q_*$ has cost $k$, and that in $Q_i$ has cost $c_{i+1}$. In $Q_T$, give each outside-component vertex cost equal to its quotient degree plus its terminal multiplicity. Its total deleted cost is exactly
\[
 |\delta_G(S_0\setminus\{z\})|+|T\setminus S_0|=b_J(S_0)=d_0.
\]
These costs dominate quotient degrees and terminal multiplicities. Cuts lift to the original graph, to $J[S_j]$, or to a child quotient, respectively, giving the following parameters:
\begin{center}
\begin{tabular}{@{}ccccc@{}}
\toprule
Quotient & Terminal mass & Expansion & Deleted cost & Remaining cost\\
\midrule
$Q_T$ & $k$ & $\alpha_0$ & $d_0$ & $M$\\
$Q_*$ & $C_*$ & $\beta$ & $k$ & $M_*\le M$\\
$Q_i$ & $m_i+c_{i+1}$ & $\eta$ & $c_{i+1}$ & $m_i$\\
\bottomrule
\end{tabular}
\end{center}
Every remaining vertex has cost less than $R$. Every graph after its prescribed deletion is a minor of $G$: its branch sets lie, respectively, in $S_0\setminus\{z\}$, $S_j\setminus\{z\}$, or $S_i\setminus S_{i+1}$. Thus any grid returned by Corollary~\ref{cor:robust-quotient-budget} is valid in $G$. If $S_j=\{z\}$, the corollary's empty-remainder convention handles $Q_*$ directly.

Suppose all these calls instead return separators. Equation~\eqref{eq:robust-quotient-budget} gives
\[
 \alpha_0k\le6d_0+t\sqrt M,\qquad
 \beta C_*\le6k+t\sqrt M,\qquad
 \eta m_i\le6C_*+t\sqrt{m_i}\quad(i<s).
\]
Since $d_0<\alpha_0k/100$, the first inequality gives $k\le2t\sqrt M/\alpha_0$. The second then gives $C_*\le13t\sqrt M/(\alpha_0\beta)$. Summing the remaining inequalities and using Cauchy--Schwarz yields
\[
 \eta M\le6sC_*+t\sqrt{sM}\le\frac{79s}{\alpha_0\beta}\,t\sqrt M.
\]
On the other hand, surviving costs in $Q_T$ dominate surviving terminal mass, so $M\ge k-d_0\ge99k/100$. Substituting $t=6Cg\sqrt{R\mathcal L}$ therefore gives
\[
 g\ge\frac{\alpha_0\beta\eta}{474C(s+1)}
       \sqrt{\frac{99k}{100R\mathcal L}}.
\]
This contradicts our choice of a sufficiently small target constant, because $\eta=\theta/2$, $s\le d$, and $\mathcal L=O(\log(2n))$. Consequently one of the quotient calls returns the required grid. There are $s+2=O(d)$ such calls, all polynomial-time.
\end{proof}

\begin{algorithm}[tb]
\caption{\textsc{AccessibleRouterOrGrid}$(G,T,R,g)$}
\label{alg:accessible-router}
\begin{algorithmic}[1]
\Require The hypotheses of Theorem~\ref{thm:accessible-router-or-grid}, with $g$ no larger than its guaranteed grid side.
\State Add the terminal hub $z$ and construct the routing hierarchy.
\State Put $B=\alpha_0|T|/100$ and $\eta=\theta/2$; let $\mathcal K$ be the component of hierarchy edges of capacity at least $B$ containing the hub leaf.
\If{a child $Y$ of a node in $\mathcal K$ has $z\notin Y$ and $R\le b_J(Y)<B$}
  \State \Return $Y$. \Comment{Lemma~\ref{lem:hierarchy-access} guarantees accessibility.}
\EndIf
\If{$\mathcal K$ contains a cluster not containing $z$}
  \State Choose a deepest such cluster $S$ and contract its children in $J[S]$.
  \State Apply Corollary~\ref{cor:robust-quotient-budget} with full child-boundary costs and terminal weights, and no prescribed deletion.
  \State \Return the lifted grid.
\EndIf
\State Write the hub chain as $S_0\supset\cdots\supset S_s=\{z\}$ and choose $S_j$ of maximum full boundary.
\State Construct the quotient instances $Q_T,Q_*,Q_0,\ldots,Q_{s-1}$, with costs, terminal weights, and prescribed deletions from the proof.
\For{each quotient instance}
  \State Run Corollary~\ref{cor:robust-quotient-budget} with target $g$.
  \If{it returns a grid}
    \State Lift the model through its surviving connected pieces and \Return the grid in $G$.
  \EndIf
\EndFor
\State \Return failure. \Comment{Excluded by the proof on promised inputs.}
\end{algorithmic}
\end{algorithm}

\FloatBarrier
\section{Retaining a large base beside the hair}\label{sec:asymmetric-split}

\begin{lemma}[An accessible side gives an unbalanced split]
\label{lem:accessible-unbalanced-split}
Fix a constant degree bound $\Delta$ and a constant
$0<\alpha_0\le1$.
Let $G$ be a connected graph on $n$ vertices of maximum degree
$\Delta$, and let $T_1,T_2$ be disjoint sets of $k$ vertices each
such that $T=T_1\cup T_2$ is fully $\alpha_0$ edge-well-linked.
Put $L=\lceil\log_2(n+2)\rceil$, and add a hub $z$ with one edge
to every terminal, obtaining $J$.

Suppose a connected set $Y\subseteq V(G)$ is supplied such that
\[
R\le b:=|\delta_J(Y)|\le\frac{\alpha_0 k}{16},
\]
its full boundary in $J$ is $\beta$ edge-well-linked in $G[Y]$,
where $\beta\ge c_\beta L^{-2}$, and its $b$ boundary incidences
can send one unit each to $z$ with edge congestion at most
$C_Q L^4$. Here $c_\beta,C_Q>0$ are fixed constants.
There are constants $c,C>0$ such that, if $R\ge CL^{10}$,
a polynomial-time algorithm finds a connected
$X\subseteq V(G)\setminus Y$ and disjoint sets
$A,B,Z\subseteq X$, together with a set $Y_0\subseteq Y$, satisfying
\begin{enumerate}
\item $A\subseteq T_1$, $B\subseteq T_2$, and
      $|A|=|B|=K\ge ck/L^2$;
\item $|Z|=|Y_0|=h\ge cR/L^{10}$, with $h\le K/2$;
\item $A,B,Z$ are individually node-well-linked in $G[X]$ and
      every pair of them is linked in $G[X]$;
\item $Y_0$ is node-well-linked in $G[Y]$; and
\item there are $h$ vertex-disjoint paths from $Z$ to $Y_0$,
      internally disjoint from $X\cup Y$.
\end{enumerate}
The constants depend only on the fixed constants in the hypotheses.
\end{lemma}

\begin{proof}
We first keep the two large interfaces. Apply \textsc{FlowTrim}
with reference $(G,T,\alpha_0)$ to
$V(G)\setminus Y$, and let $X$ be the returned terminal component.
Since $|T\cap Y|\le b$ and $|\delta_G(Y)|\le b$,
Lemma~\ref{lem:flow-trimming} gives
\[
|T_i\cap X|\ge k-b-\frac{2b}{\alpha_0}
\qquad(i=1,2).
\]
In particular the output is nonempty, $|T\cap X|\ge|T|/2$,
and the same lemma makes
$T_X\cup\Gamma_G(X)$ fully $\alpha_0/7$ edge-well-linked,
where $T_X=T\cap X$.

\paragraph{Moving the side's access into $X$.}
Write $Q=C_Q L^4$, and let $\sigma(v)$ count the boundary incidences
of $Y$ at $v$; thus $\sigma(V(G))=b$ and $\sigma(v)\le\Delta+1$.
The access guarantee gives, for every $U\subseteq V(G)$,
\[
 \sigma(U)\le Q\bigl(|\delta_G(U)|+|T\cap U|\bigr).
\]
At most half of $T$ was lost in forming $T_X$, so
\[
 |T\cap U|-2|T_X\cap U|
 \le\min\{|T\cap U|,|T\setminus U|\}
 \le |\delta_G(U)|/\alpha_0.
\]
Combining the two inequalities yields
\[
 \sigma(U)-2Q|T_X\cap U|
 \le Q(1+1/\alpha_0)|\delta_G(U)|.
\]
These are precisely the cut inequalities for sending the supply
$\sigma$ to $T_X$, with edge capacity $Q(1+1/\alpha_0)$ and sink
capacity $2Q$ at each terminal. Hence such a flow exists.
The cut $\delta_J(Y)$ implies $Q\ge1$. Put $E=Q(1+1/\alpha_0)$.
After cancelling antiparallel flow, the load at any vertex is at most
$\Delta E+\Delta+1\le2(\Delta+1)E$: the incoming edge flow contributes
at most $\Delta E$, and its own supply contributes at most $\Delta+1$.
Divide the flow by $2(\Delta+1)E$ to make every vertex load at most one.
An integral maximum flow with
unit vertex capacities consequently gives
\[
 r=\Omega(b/L^4)\ge\Omega(R/L^4)
\]
vertex-disjoint paths from $\Gamma_J(Y)$ to $T_X$.
The algorithm computes this unit-capacity flow directly; the
inequalities above are its feasibility proof.

Retain on each path the segment from its last vertex in $Y$ to its
first subsequent vertex in $X$. These segments remain disjoint and
internally avoid both regions. Their endpoints lie in
$\Gamma_G(Y)$ and $\Gamma_G(X)$, respectively.

\paragraph{Boosting the interfaces together.}
Set $\beta_0=\min\{1,c_\beta L^{-2}\}$. The $Y$ endpoints
inherit full $\beta_0$ edge-well-linkedness.
The constructive boosting theorem~\cite[Theorem~2.14]{CC16}
therefore retains a node-well-linked subset of
$\Omega(r\beta_0/L)=\Omega(R/L^7)$ endpoints.
Keep their matched $X$ endpoints, denoted by $Z_0$.
Remove $Z_0$ from $T_1\cap X$ and $T_2\cap X$.
Because $|Z_0|\le r\le b$, each remaining class has size at least
\[
k-2b-\frac{2b}{\alpha_0}\ge\frac{3k}{4}.
\]
These two classes and $Z_0$ are disjoint subsets of the fully
$\alpha_0/7$ well-linked set $T_X\cup\Gamma_G(X)$.

Apply the constructive partitioned boosting theorem
of Chekuri and Chuzhoy~\cite[Corollary~2.15]{CC16} to these
three classes. It retains a constant times $1/L$ of each class,
with every retained class node-well-linked and every pair linked.
Consequently the first two classes have size $\Omega(k/L)$,
and the retained subset of $Z_0$ has size $\Omega(R/L^8)$.
The corresponding $Y$ endpoints remain node-well-linked,
being a subset of the previously boosted set.

Thin the first two classes to a common size, and thin the
matching if needed to make $h\le K/2$.
Since $R\le b\le k$, sufficiently small $c$ and sufficiently
large $C$ leave $K\ge ck/L^2$ and $h\ge cR/L^{10}$
after all integer rounding. Thinning preserves all
node-well-linkedness and linkedness properties.
The retained paths give the required matching
between $Z$ and $Y_0$.
All steps use maximum flow or the cited polynomial boosting
algorithms.
\end{proof}

\paragraph{Using the split locally.}
Retain $G[X]$ and the paths from $Z$ to $Y_0$, but no edges
inside $Y$. The vertices of $Y_0$ are leaves in this local graph.
Since $A,B,Z$ are pairwise linked, maximum flow supplies the
$A$--$B$ linkage of size $K$ and the $Y_0$--$A$ linkage of
size $h$ required by Theorem~\ref{thm:unbalanced-local}.
The unused edges of $G[Y]$ remain available to realize matchings
between the selected hair endpoints during the final assembly.

\paragraph{Algorithm for Lemma~\ref{lem:accessible-unbalanced-split}.}
\begin{enumerate}
\item Compute $X=\textsc{FlowTrim}(G,T,\alpha_0,V(G)\setminus Y)$ and put $T_X=T\cap X$.
\item In $G$, compute a maximum collection of vertex-disjoint paths from $\Gamma_J(Y)$ to $T_X$, with unit vertex capacities also at the sources and sinks. The cut inequalities in the proof guarantee $r=\Omega(b/L^4)$ paths; no auxiliary routing is computed.
\item On each path retain the segment from its last vertex in $Y$ to its first subsequent vertex in $X$. Let $P_Y$ and $P_X$ be the two endpoint sets and retain their path matching.
\item Set $\beta_0=\min\{1,c_\beta L^{-2}\}$ and apply the constructive boosting algorithm of \cite[Theorem~2.14]{CC16} with parameter $\beta_0$ to $P_Y$ in $G[Y]$. Retain its node-well-linked output and the corresponding matched endpoints $Z_0\subseteq P_X$.
\item Form the three disjoint classes $(T_1\cap X)\setminus Z_0$, $(T_2\cap X)\setminus Z_0$, and $Z_0$. Apply \cite[Corollary~2.15]{CC16} with parameter $\alpha_0/7$ inside $G[X]$, obtaining retained classes $A',B',Z'$.
\item Set $K=\min\{|A'|,|B'|\}$ and $h=\min\{|Z'|,\lfloor K/2\rfloor\}$. Choose arbitrary $K$-subsets $A\subseteq A'$ and $B\subseteq B'$, and an arbitrary $h$-subset $Z\subseteq Z'$. Let $Y_0$ be its matched endpoints in $Y$, and retain the corresponding $h$ paths.
\item Return $X,A,B,Z,Y_0$ and these paths.
\end{enumerate}
\section{Certified crossbar assembly}
\label{sec:polynomial-assembly}

The assembly uses the original strong path-of-sets system throughout: each odd cluster supplies one matching, and each even cluster joins consecutive crossbars. There is no need to build a second strong system after splitting the odd clusters. To obtain a polynomial-time cut player, we use the random-projection argument of Khandekar, Rao, and Vazirani~\cite{KRV09} with an exact final test. Its $O(\log^2 N)$ rounds replace the logarithmic-round cut player used in the structural assembly of~\cite{CT21,KKOV07}, changing only logarithmic factors.

\FloatBarrier

\begin{lemma}[An exact polynomial cut player]
\label{lem:exact-cut-player}
Let $N\ge2$ be even, put $s=\lceil\log_2(2N)\rceil$, and let $r=128s^2$. There is a randomized polynomial-time cut player for $r$ rounds on $[N]$ with the following guarantee. If every requested bisection receives a perfect matching, then, with probability at least $7/8$, the algorithm returns their union $J$ and a polynomially checkable certificate that
\[
 |E_J(U,[N]\setminus U)|\ge |U|/2
 \qquad (|U|\le N/2).
\]
The matching player may depend on the full history and the current random choices. The graph $J$ retains edge multiplicities and is $r$-regular. Every returned certificate is valid; an invalid matching or a failed final test causes the algorithm to return failure. The stated running time excludes the work of supplying the matchings.
\end{lemma}

\paragraph{Algorithm \textnormal{\textsc{CertifiedCutMatching}}\((N,\mathsf{Match})\).}
\begin{enumerate}
\item Set $r=128\lceil\log_2(2N)\rceil^2$ and $A=I_N$.
\item For $j=1,\ldots,r$:
  \begin{enumerate}
  \item Choose fresh independent uniform signs $\xi\in\{-1,1\}^N$. Sort the coordinates of $A\xi$, breaking ties by vertex labels, and let $(L_j,R_j)$ be the first and last $N/2$ labels.
  \item Obtain $M_j=\mathsf{Match}(L_j,R_j)$. Return failure unless it is a perfect matching across this bisection.
  \item Simultaneously replace the two rows of $A$ indexed by each edge of $M_j$ by the sum of their old rows. Store $M_j$.
  \end{enumerate}
\item Return failure unless
\[
 \sum_{i,v}(NA_{iv}-2^r)^2\le 2^{2r-2}.
\]
Otherwise return $J=M_1\cup\cdots\cup M_r$ and the matching list.
\end{enumerate}

\begin{proof}
For the permutation matrix $\Pi_j$ of $M_j$, put $D_j=(I_N+\Pi_j)/2$ and $P_j=D_jP_{j-1}$, starting from $P_0=I_N$. The algorithm maintains $A=2^jP_j$. Each $P_j$ is doubly stochastic. Write $p_i$ for its rows, $\bar p=(1/N,\ldots,1/N)$, and
\[
 \Psi(P)=\sum_i\|p_i-\bar p\|_2^2.
\]
Averaging a matched pair gives the exact identity
\begin{equation}\label{eq:exact-potential-drop}
 \Psi(P)-\Psi(P')=\frac12\sum_{\{i,j\}\in M}\|p_i-p_j\|_2^2.
\end{equation}

\paragraph{Expected progress.}
Fix the history before one round. For its fresh signs, let $u_i=\langle p_i-\bar p,\xi\rangle$ and $S=\sum_i u_i^2$. Then $\sum_i u_i=0$, $\mathbb E S=\Psi(P)$, and $S\le N\Psi(P)$. The random-sign tail bound, followed by a union bound over row pairs, shows that the event
\[
 \mathcal E=\left\{
 (u_i-u_j)^2\le8\ln(2N)\|p_i-p_j\|_2^2
 \text{ for all }i,j
 \right\}
\]
satisfies
\begin{equation}\label{eq:all-pairs-event}
 \Pr(\mathcal E^c)\le N^2\cdot2e^{-4\ln(2N)}
 =\frac1{8N^2}.
\end{equation}
The simultaneous bound is what permits an adaptive matching player. For any median $m$ between the two sorted halves, every matching across them satisfies
\[
 \sum_{\{i,j\}\in M}(u_i-u_j)^2
 \ge\sum_i(u_i-m)^2=S+Nm^2\ge S.
\]
Thus on $\mathcal E$ the potential drop is at least $S/(16\ln(2N))$. It is always nonnegative, so, conditional on the fixed history,
\[
 \mathbb E[\Psi(P)-\Psi(P')]
 \ge\frac{\mathbb E S-N\Psi(P)\Pr(\mathcal E^c)}{16\ln(2N)}
 \ge\frac{\Psi(P)}{32\ln(2N)}.
\]
Since $\Psi(P_0)=N-1\le N$ and $s\ge\ln(2N)$, iteration gives
\[
 \mathbb E\Psi(P_r)
 \le N e^{-r/(32\ln(2N))}\le\frac1{16N^3}.
\]
Consequently
\begin{equation}\label{eq:cut-player-success}
 \Pr\!\left[\Psi(P_r)>\frac1{4N^2}\right]
 \le\frac1{4N}\le\frac18.
\end{equation}

\paragraph{The certificate.}
The integer test in the algorithm is exactly $\Psi(P_r)\le1/(4N^2)$, which implies $(P_r)_{iv}\ge1/(2N)$ for every $i,v$. For $|U|\le N/2$, start with one unit of mass at each vertex of $U$, and apply the lazy matchings in reverse order, so that their transition matrix is $P_r=D_r\cdots D_1$. Each vertex always has mass between zero and one. A crossing matching edge can increase the mass outside $U$ by at most $1/2$. Hence
\[
 \frac12|E_J(U,[N]\setminus U)|
 \ge \sum_{i\in U,\ v\notin U}(P_r)_{iv}
 \ge\frac{|U|(N-|U|)}{2N}\ge\frac{|U|}{4}.
\]
This proves the claimed expansion for every transcript that passes the test.

The verifier checks that the list contains exactly $r$ matchings and that every supplied $M_j$ is a perfect matching of $[N]$, recomputes $A$ from them beginning with $I_N$, and checks the same integer inequality. After round $j$, every entry of $A$ lies in $[0,2^j]$. There are $O(rN^2)$ arithmetic operations and comparisons on integers represented with $O(r+\log N)$ digits, so both the algorithm and verifier run in polynomial time.
\end{proof}

\begin{proposition}[Polynomial assembly from the original chain]
\label{prop:polynomial-crossbar-assembly}
For even $N\ge2$, define
\begin{equation}\label{eq:assembly-parameters}
 s_N=\lceil\log_2(2N)\rceil,\qquad r_N=128s_N^2,\qquad
 h_N=\max\{t\in\mathbb Z_{\ge0}:384^2r_N^2s_Nt^2\le N\}.
\end{equation}
Write $r=r_N$ and $h=h_N$, and assume $h\ge2$. Suppose that $G$ is supplied with a strong path-of-sets system $C_1,\ldots,C_{2r-1}$ of width at least $N$, with original interfaces $(A_i,B_i)$. In every odd cluster $C_i$, suppose we are also given:
\begin{enumerate}
\item a crossbar with $N$ disjoint backbones whose left and right endpoints belong to $A_i$ and $B_i$, respectively, and $N$ disjoint attachment paths, one from each backbone, ending at a set $X_i$; and
\item a connected hair subgraph $S_i\subseteq C_i$ in which $X_i$ is node-well-linked, such that $S_i$ is disjoint from the backbones and meets the attachment paths only at $X_i$.
\end{enumerate}
All subgraphs and paths are supplied explicitly. There is a randomized algorithm, polynomial in their total size, that returns a verified $h\times h$ grid-minor model with probability at least $7/8$, and otherwise returns failure.
\end{proposition}

\begin{proof}
\emph{Stitch the backbones.}
Keep only the original connector paths incident with the $N$ chosen backbone endpoints at each odd cluster. Between consecutive odd clusters, their opposite endpoints form $N$-subsets of the two interfaces of the intervening even cluster. Linkedness supplies a linkage between these subsets; compute one by unit-vertex-capacity maximum flow. No prescribed pairing is needed. Concatenating the crossbar backbones, selected connectors, and these buffer linkages gives $N$ disjoint global backbones $R_1,\ldots,R_N$.

Only the selected connectors are used. Their interiors avoid every original cluster, and their odd endpoints lie on crossbar backbones, outside the corresponding hair. Thus no retained connector enters a hair, even when some unselected original interface vertices lie in that hair. No thinning to a common new interface width is required.

\paragraph{Supply the matchings.}
Each of the $r$ odd clusters has one attachment from every global backbone to its hair. At round $j$ of Lemma~\ref{lem:exact-cut-player}, partition these $N$ hair endpoints by the requested bisection of the backbone labels. Node-well-linkedness gives $N/2$ disjoint paths between the halves, found by another unit-vertex-capacity maximum flow. Together with the two attachments at their ends, these paths realize a perfect matching across the bisection.

The matching-edge paths have disjoint interiors, both within a round and across the disjoint original odd clusters, and their interiors avoid all global backbones. They therefore represent the multigraph $J$ explicitly as a minor of $G$. For an ordinary branch-set model, orient each edge path, add its internal vertices to the backbone branch set at its initial end, and use its last edge as the adjacency witness. If the cut player's final test fails, return failure. Otherwise $J$ is a certified $1/2$-edge-expander of degree $r$.

\paragraph{Extract and lift the grid.}
Put $a=1/(48r)$ and let $J_0$ be the underlying simple graph of $J$. There is no vertex separator $Z$ of size at most $aN$ whose deletion leaves components of size at most $2N/3$. Indeed, grouping components gives a union $U$ with $N/6\le|U|\le N/2$, and every edge leaving $U$ meets $Z$. Expansion in the multigraph then gives the contradiction
\[
 N/12\le|U|/2\le|E_J(U,[N]\setminus U)|
 \le r|Z|\le N/48.
\]
Apply Lemma~\ref{lem:weighted-grid-primitive} to $J_0$ with unit costs and separator parameter $a$. Its hypothesis holds because
\[
 h^2\le\frac{N}{384^2r^2s_N}
       =\frac{a^2N}{64s_N}.
\]
The separator outcome is impossible, so the primitive returns an $h\times h$ grid model. This use of the primitive requires no relation between costs and degrees.

Compose that model with the stored branch sets and edge witnesses in $G$, choosing one witness for each edge of $J_0$. Each resulting branch set is connected, and the branch sets remain disjoint. Contraction forests allow this composition without repeated expansion; the final branch sets together contain at most $|V(G)|$ vertices. Check the model directly in $G$ before returning it.

There are $r-1$ buffer flows and $r$ hair flows, the polynomial cut-player computation, and one call to the constructive grid primitive. All steps are polynomial. On valid supplied data, the only possible failure is the cut player's final test, which fails with probability at most $1/8$.
\end{proof}

\paragraph{Larger targets and failed suppliers.}
If $h\ge g$, restrict the returned model to a $g\times g$ subgrid. In the general-graph algorithm, check supplied paths, sizes, and disjointness explicitly, and place every routine under its fixed polynomial clock; no exact well-linkedness test is used. An unsuccessful flow, invalid explicit witness, or timeout returns failure. Every grid model, including an earlier exit, is lifted and checked in the original graph. Thus failed supplier promises cannot cause a false positive, while on the good supplier events the stated clocks do not alter the success probability.

\section{Reducing the exponent through colorful coverings}
\label{sec:conditional-cover}

The sixth-power proof turns a failed crossbar search into a strong path-of-sets system. A sufficiently strong packing theorem would supply the crossbar directly. We formulate the resulting exponent tradeoff uniformly. Colorful-minor packing is studied by Protopapas, Thilikos, and Wiederrecht~\cite{PTW26}; the hypothesis below asks for a quantitative bound in graphs excluding the target grid.

For disjoint terminal sets $A,B,X\subseteq V(H)$, call a connected subgraph \emph{colorful} if it meets all three sets. Let $\nu(H;A,B,X)$ be the maximum number of vertex-disjoint colorful subgraphs, and let $\tau(H;A,B,X)$ be the minimum size of a vertex set meeting every colorful subgraph. The two parameters are unchanged if we restrict the colorful subgraphs to trees.

\begin{definition}[Colorful covering hypothesis at exponent $a$]\label{hyp:colorful-cover}
For a fixed real $a\ge0$, let $\mathsf C(a)$ be the following assertion. There are constants $C\ge1$ and $c\ge0$ such that, for every integer $g\ge2$, every graph $H$ with no $g\times g$ grid minor, and every three disjoint terminal sets $A,B,X$,
\[
 \tau(H;A,B,X)\le Cg^a\nu(H;A,B,X)
       \bigl(\log_2(g\nu(H;A,B,X)+2)\bigr)^c.
\]
\end{definition}

The exponent $a$ is fixed before the input is chosen. Each value gives a separate hypothesis; establishing any $\mathsf C(a)$ with $a<4$ would improve the sixth-power bound by the theorem below. The case $a=1$ gives exponent three, matching the exponent in the classical cubic conjecture up to logarithmic factors. A two-terminal path-packing theorem does not imply this three-terminal covering hypothesis.

\begin{lemma}[Two linkages force a large colorful transversal]
\label{lem:colorful-linkages}
Let $A,B,X$ be disjoint sets of common size $\kappa$. If there are full linkages from $A$ to $B$ and from $A$ to $X$, then
$\tau(H;A,B,X)\ge\kappa/2$.
If every vertex of $X$ has degree one, a packing of $\rho$ colorful subgraphs contains an $(A,B,X)$-crossbar of width $\rho$.
\end{lemma}
\begin{proof}
For each $a\in A$, the two linkage paths starting at $a$ contain a colorful tree $T_a$. Each vertex belongs to at most two of these trees, one for each linkage family. A transversal meeting all $\kappa$ trees therefore has size at least $\kappa/2$.

In each member of a colorful packing, retain a minimal tree joining chosen vertices $a\in A$, $b\in B$, and $x\in X$. Its $a$--$b$ path and the path from $x$ to its first contact with that path form one crossbar component. The latter path is nontrivial because $x$ is a leaf of $H$. Disjoint colorful trees give disjoint components.
\end{proof}

\begin{theorem}[Conditional exponent transfer]\label{thm:colorful-transfer}
For every fixed $a\ge0$, hypothesis $\mathsf C(a)$ implies
\[
 f(g)=\widetilde O(g^{a+2}).
\]
Thus every decrease in the covering exponent gives the same decrease in the resulting excluded-grid exponent.
\end{theorem}
\begin{proof}
Put $L=\log_2(g+2)$, and let $q$ be the least power of two at least $A_0gL^{3/2}$, where $A_0$ is a sufficiently large absolute constant. Set
\[
 \rho=q^2,\qquad
 \kappa=\left\lceil4Cg^a\rho
       \bigl(\log_2(g\rho+2)\bigr)^c\right\rceil .
\]
Here $C,c$ come from $\mathsf C(a)$. Since $a,c$ are fixed,
\[
 \rho=O(g^2L^3),\qquad
 \kappa=O(g^{a+2}L^{c+3}).
\]

Suppose $G$ excludes the $g\times g$ grid. The hairy-system supplier~\cite[Theorem~2.3]{CT21} gives a hairy path-of-sets system of width $\kappa$ and length $\ell=\Theta(\log q)$ whenever $\tw(G)\ge\kappa L^{O(1)}$. Choose $\ell$ even and large enough for the crossbar-chain theorem~\cite[Theorem~3.2]{CT21}. In each required odd cluster, the local graph has a full $A_i$--$B_i$ linkage and a full $A_i$--$Y_i$ linkage, where $Y_i$ consists of degree-one hair endpoints. These graphs also exclude the target grid.

Let $\nu_i,\tau_i$ be their colorful packing and covering numbers. If $\nu_i<\rho$, the preceding lemma and $\mathsf C(a)$ give
\[
 \frac{\kappa}{2}\le\tau_i
 < Cg^a\rho\bigl(\log_2(g\rho+2)\bigr)^c
 \le\frac{\kappa}{4},
\]
a contradiction. Each required cluster therefore contains a crossbar of width $q^2$. The crossbar-chain theorem gives a grid of side $\Omega(q/(\log q)^{3/2})\ge g$, again a contradiction. The covering hypothesis uses the excluded side $g$ throughout; $q$ only absorbs the assembly's logarithmic loss.

To verify the supplier threshold explicitly, write its condition as
$k/(\log_2 k)^{c_0}>c_1\kappa\ell^{48}$.
For a sufficiently large constant $C_0$, take
\[
 K(g)=\left\lceil C_0\kappa L^{48+c_0}\right\rceil.
\]
Then $\log K(g)=O(L)$ and the supplier condition holds with $k=K(g)$. Hence
$f(g)\le K(g)=O(g^{a+2}L^{c+51+c_0})$, as claimed.
\end{proof}

This transfer is existential. A covering inequality alone does not provide an algorithm for the colorful packing, and the proof also uses the existential hairy-system supplier. A polynomial-time result at a smaller exponent would require constructive versions of these ingredients at the corresponding parameters. Once the packing is available, its conversion to a crossbar uses only tree pruning and path extraction.

\section{Parameter choices and the general-graph construction}
\label{sec:parameters}

We collect the local and global budgets in this section. The logarithmic powers are deliberately generous; each constraint is checked after all relevant constants have been fixed.

\subsection{Local parameters}

This section fixes the parameters deferred by the proof. We use generous logarithmic powers so that every constraint can be checked in one place. The construction never pays an additional polynomial factor in $w$, $\rho$, or $\ell$.

\begin{proposition}[Constructive pseudo-grid extraction]\label{prop:consumer}
There is an absolute constant $A_{\rm loc}\ge2$ with the following property. For integers $w,\rho,\ell\ge2$, set
\begin{equation}\label{eq:explicit-budgets}
 L=\lceil\log_2(w\rho\ell+2)\rceil,
 \qquad H=\lceil A_{\rm loc}wL^{40}\rceil,
 \qquad m=\lceil A_{\rm loc}^2w\rho\ell L^{50}\rceil.
\end{equation}
Given a maximum-degree-four host covered by $N\le64H\rho$ perfect unique rows and $m$ disjoint auxiliary paths each meeting $64H$ distinct rows, one can find in polynomial time a strong path-of-sets system of width $w$ and length $\ell$.
\end{proposition}
\begin{proof}
Write $h=\lceil\log_2(32m+2)\rceil$ and take the common approximation bound $r=r(16m)=O(h)$. Increasing $A_{\rm loc}$ makes $m\ge N$. Apply Lemma~\ref{lem:source}, with ceilings in the logarithmic denominator of $\alpha_0$. The total terminal mass satisfies $T_0\in[m,16m]$, the minimum cluster mass is $8H$, and
\[
 \alpha_0=\Omega(h^{-2}),\qquad
 d=\lceil\log_2(T_0+1)\rceil+2=O(h).
\]
Take $\Delta=4$ and set
\begin{equation}\label{eq:local-parameters}
 \delta=\frac{\alpha_0}{64\Delta d},\qquad
 \beta=\frac{\alpha_0}{8},\qquad
 \gamma=\frac{\beta}{100r\lceil\log_2(2N+2)\rceil},\qquad
 W=\lfloor\delta H/16\rfloor.
\end{equation}
Then $128W/\delta\le8H$, so localization applies. Its losses and the final width are summarized below. Each hidden constant is absolute.

\begin{center}
\begin{tabular}{@{}lll@{}}
\toprule
Quantity & Bound & Reason\\
\midrule
Capture fraction $\delta$ & $\Omega(h^{-3})$ & $\alpha_0/(64\Delta d)$\\
Port well-linkedness $\beta$ & $\Omega(h^{-2})$ & export from a trimmed cluster\\
Endpoint well-linkedness $\gamma$ & $\Omega(h^{-4})$ & one approximate decomposition\\
Per-cluster scale $W$ & $\Omega(H/h^3)$ & choice in~\eqref{eq:local-parameters}\\
Total assignment mass & $\Omega(m/h^4)$ & Corollary~\ref{cor:ordered-mass}\\
Weak width $u$ & $\Omega(H/h^7)$ & Lemma~\ref{lem:ordered}\\
Strong width & $\Omega(H/h^{21})$ & $\Omega(\gamma^4W/r^2)$\\
\bottomrule
\end{tabular}
\end{center}

Since $N\le64H\rho$, sufficiently large absolute constants $C_1,C_2$ make the following two inequalities sufficient for the routing mass condition, all integer rounding requirements, and final width at least $w$:
\begin{equation}\label{eq:two-constraints}
 m\ge C_1H\rho\ell h^4,
 \qquad H\ge C_2wh^{21}.
\end{equation}
The first ensures $\delta T_0/(8d)>2N(\ell-1)$, with slack. The second ensures $W\ge1$, $u\ge1$, and that all boosting and thinning widths exceed their requested integer values.

For~\eqref{eq:explicit-budgets}, there is an absolute $C_3$ such that
\[
 h\le C_3(1+\log_2 A_{\rm loc})L.
\]Thus~\eqref{eq:two-constraints} follows after enlarging the constants from
\[
 A_{\rm loc}\ge C_4(1+\log_2 A_{\rm loc})^4,
 \qquad
 A_{\rm loc}\ge C_5(1+\log_2 A_{\rm loc})^{21}.
\]
Fix one absolute $A_{\rm loc}$ satisfying these and the finitely many constant-size rounding requirements. Such an $A_{\rm loc}$ exists because $A_{\rm loc}/(1+\log_2 A_{\rm loc})^{21}$ tends to infinity. The powers $L^{40}$ and $L^{50}$ leave more than the required logarithmic slack. Lemmas~\ref{lem:source}, \ref{lem:ordered}, and \ref{lem:strong}, together with Corollary~\ref{cor:ordered-mass}, now prove the proposition constructively.
\end{proof}

\subsection{Global parameters}
\label{sec:global-algorithm}

All constants in the local construction, including $A_{\rm loc}$ and the resulting constants in Theorem~\ref{thm:unbalanced-local}, are fixed. We now choose a new absolute integer $A$, after the constants in the two constructive suppliers have been fixed, and finally choose an absolute power-of-two cutoff $q_0$. The supplier exponents $a_s,a_p$ may be rounded upward to integers.

For a power of two $q\ge q_0$, put
\begin{equation}\label{eq:global-parameters}
\begin{aligned}
 L&=\lceil\log_2(q+2)\rceil,&N&=q^2,&
 s&=\lceil\log_2(2N)\rceil,&r&=128s^2,\\
 L_0&=2r-1,&
 R&=\lceil A q^4L^{60}\rceil,&
 k&=\lceil A^2q^6L^{80}\rceil,\\
 D_0&=180+a_s+a_p,&
 F&=\lceil A^{10}q^6L^{D_0}\rceil,&
 \Lambda&=\lceil\log_2(C_sF^4+3)\rceil .
\end{aligned}
\end{equation}
Here $k$ is the width of the supplied strong chain, and $R$ is the smaller boundary scale. The parameter $F$ is the promised treewidth threshold used in the sparsifier. All are chosen before invoking it.

There are fixed constants $C_1,\ldots,C_5$ such that the following conditions suffice:
\begin{align}
 F&\ge C_1 kL_0^{48}\Lambda^{a_s+a_p},\label{eq:global-supplier-budget}\\
 k&\ge C_2 q^2R\Lambda^{13},\label{eq:global-router-budget}\\
 k/\Lambda^2&\ge C_3 q^6L^{50},\label{eq:global-base-budget}\\
 R/\Lambda^{10}&\ge C_4 q^4L^{40},\label{eq:global-hair-budget}\\
 R&\ge C_5\Lambda^{10},\qquad R>4,\qquad R<k/100.\label{eq:global-size-budget}
\end{align}
The first inequality pays for the sparsifier's treewidth loss and the strong path-of-sets supplier. In a cluster of order at most $C_sF^4$, the hierarchy has $d=O(\Lambda)$, $\beta=\Omega(\Lambda^{-2})$, and $\theta=\Omega(\Lambda^{-3})$. Thus Theorem~\ref{thm:accessible-router-or-grid} guarantees a grid of side
\[
 \Omega\!\left(\sqrt{k/(R\Lambda^{13})}\right).
\]
The second inequality makes that side at least $q$. The third and fourth pay for the unbalanced local theorem after splitting: its input has base width $\Omega(k/\Lambda^2)$ and hair width $\Omega(R/\Lambda^{10})$. Its local logarithm is at most $6L$, which is absorbed in $C_3,C_4$. The last line gives the hypotheses of the hierarchy and split routines.

We verify these requirements together. There is a fixed $C_D$ such that
\[
 \Lambda\le C_D(1+\log_2 A)L.
\]
Also $s\le3L$, $r\le1152L^2$, and $L_0\le2304L^2$. Hence the right side of~\eqref{eq:global-supplier-budget} is at most a fixed constant times
\[
 A^2(1+\log A)^{a_s+a_p}
 q^6L^{176+a_s+a_p},
\]
whereas $F=A^{10}q^6L^{180+a_s+a_p}$ up to rounding. The hierarchy requirement compares $A^2q^6L^{80}$ with a fixed constant times $Aq^6L^{73}(1+\log A)^{13}$. The base and hair requirements reduce to
\[
 A^2L^{28}\ge C(1+\log A)^2,\qquad
 AL^{10}\ge C(1+\log A)^{10},
\]
respectively. The remaining size conditions follow by increasing the same fixed constants. Choose an integer $A$ large enough to satisfy the finitely many inequalities
\[
 A^8\ge C(1+\log_2 A)^{a_s+a_p},\qquad
 A\ge C(1+\log_2 A)^{13},\qquad
 A^2\ge C(1+\log_2 A)^2,
\]
with $C$ dominating the constants above. Such an $A$ exists. Fix it, and then fix $q_0$ large enough for the stated supplier probabilities and all rounding conditions. These choices depend on no input parameter.

\FloatBarrier
\subsection{The general-graph algorithm}

\begin{algorithm}[tb]
\caption{\textsc{SixthPowerGrid}$(G,g)$}
\label{alg:global-grid}
\begin{algorithmic}[1]
\Require A finite graph $G$ and an integer $g\ge2$.
\If{$g^2>|V(G)|$} \State \Return failure. \EndIf
\State Choose the least power of two $q\ge q_0$ satisfying $384^2r_{q^2}^2s_{q^2}g^2\le q^2$, using the definitions of $r_N,s_N$ in Proposition~\ref{prop:polynomial-crossbar-assembly}.
\State Set all parameters by~\eqref{eq:global-parameters}.
\State Apply the degree-three sparsifier with promised threshold $F$; retain its topological-minor map. If it fails, \Return failure.
\State Apply the strong path-of-sets supplier with width $k$ and length $L_0$.
\If{the supplier fails or its explicit data are invalid} \State \Return failure. \EndIf
\For{each odd cluster $C_i$ with original interfaces $T_{i,1},T_{i,2}$}
  \State Apply \textsc{AccessibleRouterOrGrid} to $H[C_i]$, $T_{i,1}\cup T_{i,2}$, and boundary scale $R$, with target side $q$.
  \If{a grid is returned}
    \State Lift, restrict to side $g$, verify in $G$, and \Return the model if valid.
  \EndIf
  \State Apply Lemma~\ref{lem:accessible-unbalanced-split} to its accessible side, obtaining $X_i,A_i,B_i,Z_i,Y_i$ and attachment paths.
  \State In $H[X_i]$ together with those paths, run \textsc{LocalDichotomy} with $w=\rho=\ell=N$ and leaf set $Y_i$.
  \If{a strong system is returned}
    \State Apply~\cite[Corollary~3.4]{CC16} with target side $g$; lift, verify in $G$, and \Return the model if valid.
  \EndIf
  \State Retain the returned crossbar and its hair. On any unsuccessful required step, \Return failure.
\EndFor
\State Apply Proposition~\ref{prop:polynomial-crossbar-assembly}; on a failed potential test, \Return failure.
\State Lift its grid, restrict to side $g$, verify in the original $G$, and \Return the model if valid; otherwise \Return failure.
\end{algorithmic}
\end{algorithm}

An implementation places every routine under its fixed polynomial clock and returns failure after any invalid explicit witness, unsuccessful flow request, or timeout. These checks involve paths, sizes, and incidences, not exact well-linkedness. Every successful exit, including an early exit, is verified in the original graph. The size guard is valid because a grid model needs $g^2$ disjoint nonempty branch sets. After that guard, $g\le\sqrt{|V(G)|}$; thus the algorithm also has polynomial running time for a binary-encoded target.

The suppliers are stated for arbitrary graphs. Equivalently, any connected-input implementation can be run on every connected component and its outputs processed in turn. A component satisfying the promise exists whenever the original graph satisfies it. Processing the components adds only a polynomial factor and requires no treewidth test.

\begin{proof}[Proof of Theorem~\ref{thm:main}]
First fix $q$ and the parameters above, and suppose $\tw(G)\ge F$. With probability at least $9/10$, the sparsifier gives a degree-three topological minor $H$ of order at most $C_sF^4$ and treewidth at least $F/(C_s\log^{a_s}(F+2))$. Condition~\eqref{eq:global-supplier-budget} permits the strong path-of-sets supplier to produce a chain of width $k$ and length $L_0$, with conditional probability at least $9/10$.

In each odd cluster the union of its two original interfaces is fully $1/2$ edge-well-linked, by the linkedness argument in Section~\ref{sec:prelim}. The hierarchy routine either returns a $q\times q$ grid or supplies $Y$ with
\[
 R\le b_J(Y)<\frac{(1/2)(2k)}{100}
              <\frac{(1/2)k}{16}.
\]
Its access congestion is $O(\Lambda^4)$, and its boundary well-linkedness is $\Omega(\Lambda^{-2})$. Lemma~\ref{lem:accessible-unbalanced-split} therefore applies. It preserves base width $\Omega(k/\Lambda^2)$ and hair width $\Omega(R/\Lambda^{10})$. By~\eqref{eq:global-base-budget}--\eqref{eq:global-hair-budget}, Theorem~\ref{thm:unbalanced-local} applies with $w=\rho=\ell=q^2$.

If a local call returns a strong system of width and length $q^2$, the square-system algorithm supplies the requested grid. Otherwise each odd cluster supplies a crossbar of width $N=q^2$. All its new base vertices, hair vertices, and attachments lie in that original cluster. Different working regions are therefore disjoint.

Each crossbar begins and ends in subsets of its original cluster interfaces. Its hair lies inside that same odd cluster, is disjoint from the crossbar backbones, and meets the attachment paths only at their hair endpoints. The original strong chain and these crossbars therefore satisfy Proposition~\ref{prop:polynomial-crossbar-assembly} directly. Its construction keeps only the selected original connectors and joins them through the untouched even clusters.

Apply Proposition~\ref{prop:polynomial-crossbar-assembly}. Its only random failure on this valid input is the certified cut player's failure, of probability at most $1/8$. A complete trial consequently succeeds with probability at least
\[
 (9/10)^2(7/8)=567/800>2/3.
\]
This calculation uses conditional probabilities and remains valid when the matching choices depend on the preceding history.

It remains to choose $q$ as a function of $g$. The integer stopping test in Algorithm~\ref{alg:global-grid} is exactly $h_{q^2}\ge g$. It implies $q\ge384g$, so
\[
 q^2\ge16g^2+10g,\qquad q^2\ge2g(g-1).
\]
Thus both the hierarchy exit and the local square-system exit return a grid of side at least $g$, with the latter algorithm applied directly to target $g$.

Conversely, a sufficiently large fixed multiple of
$g(\log_2(g+2))^{5/2}$, rounded upward to a power of two, satisfies the stopping test: its logarithm is $O(\log(g+2))$, and the fixed multiplier can absorb this logarithmic constant since $B/(1+\log B)^{5/2}\to\infty$. Therefore
\[
 q=O(g\log^{5/2}(g+2)),\qquad
 F=O\!\left(g^6\log^{195+a_s+a_p}(g+2)\right).
\]
Parameter selection itself uses only integer doubling and comparisons.

Every deterministic subroutine is polynomial in the explicit working graph and parameters, and the two published suppliers have polynomial running times. The working graph has order at most $C_sF^4$, while all parameters are bounded by fixed polynomials in $g$. Therefore the total running time is $(|V(G)|+|E(G)|+g)^{O(1)}$ with an absolute exponent. Contraction forests and stored edge witnesses compose the minor models in polynomial time; the final branch sets have at most $|V(G)|$ original vertices in total. Direct verification prevents a false positive on every input. Independent trials have expected constant repetition count on promised inputs, completing the theorem.
\end{proof}

\section{The lower bound and the remaining gap}
\label{sec:gap}

The classical lower bound of Robertson, Seymour, and Thomas~\cite{RST94} has a different source from the losses in our proof. A $g\times g$ grid contains $\lfloor g/2\rfloor^2$ vertex-disjoint cycles. Any minor model of it lifts these to that many disjoint cycles in the host. In a graph of girth $h$, they require at least $h\lfloor g/2\rfloor^2$ vertices. Constant-degree graphs with linear treewidth and logarithmic girth therefore give the $\Omega(g^2\log g)$ obstruction. This is the established lower bound, not a new lower-bound contribution of the present paper.

The resulting picture is
\[
 \Omega(g^2\log g)\ \le\ f(g)\ \le\ O(g^6\polylog g).
\]
The upper and lower bounds do not match. In particular, the sixth-power theorem does not settle the near-quadratic prediction, nor the cubic conjecture. The colorful covering hypothesis gives a general route to a smaller exponent: a factor $g^a$, up to logarithms, in the packing--covering ratio would give grid exponent $a+2$. The linear case would give exponent three, up to logarithmic factors.

For the connection to bidimensionality, let $p$ be a minor-monotone graph parameter satisfying $p(\Grid_g)\ge c_pg^2$ for a fixed $c_p>0$. Taking the first grid size whose parameter value exceeds $p(G)$ and applying Theorem~\ref{thm:main} gives the general structural bound $\tw(G)=\widetilde O((p(G)+1)^3)$. This is the corresponding improvement in the unrestricted parameter--treewidth implication. It does not replace the stronger, already linear treewidth/grid relation in fixed-minor-free classes, nor by itself supply a faster algorithm for every such parameter.

\paragraph{AI Use Statement.}
This work resulted from extensive interaction between the author and OpenAI Astra, with the author providing suggestions and critically checking Astra's responses through iterative exchanges. OpenAI Astra was also used in writing and revising the manuscript through detailed prompts from the author, with Gemini Argon assisting in some parts. The author directed the process and evaluated the proposed ideas and arguments. 
The public GitHub link will be added here: \texttt{https://github.com/hajiaghayi/Grid-Minor}.


\end{document}